\documentclass{aa} 
\usepackage{graphicx}
\usepackage{txfonts}
\begin{document} 

   \title{Ca II Triplet Spectroscopy of Small Magellanic Cloud Red Giants. V. Abundances and Velocities for 12 Massive Clusters}
    \titlerunning{CaT spectroscopy of SMC star cluster red giant stars.}
    
   \author{M.C. Parisi\inst{1,2,3}
          \and
          L.V. Gramajo\inst{1,3}
          \and 
          D. Geisler\inst{4,5,6}
          \and 
          B. Dias\inst{7}
          \and 
          J.J. Clari\'a\inst{1,3}
          \and 
          G. Da Costa\inst{8}
          \and
          E. K. Grebel\inst{9}
          }

\institute{
  	Observatorio Astron\'omico, Universidad Nacional de C\'ordoba, Laprida 854, X5000BGR, C\'ordoba, Argentina\\
  	    \email{cparisi@unc.edu.ar}
  	    \and
  	     Instituto de Astronom{\'\i}a Te\'orica y Experimental (CONICET-UNC), Laprida 854, X5000BGR, C\'ordoba, Argentina
  	 \and
  	    Consejo Nacional de Investigaciones Cient\'ificas y T\'ecnicas (CONICET), Godoy Cruz 2290, Buenos Aires, CPC 1425FQB, Argentina
  	 \and   
         Departamento de Astronom\'ia, Casilla 160-C, Universidad de Concepci\'on, Chile
         \and
   	    Instituto de Investigación Multidisciplinario en Ciencia y Tecnología, Universidad de La Serena. Avenida Raúl Bitrán S/N, La Serena, Chile
         \and
        Departamento de Astronomía, Facultad de Ciencias, Universidad de La Serena, Av. Juan Cisternas 1200, La Serena, Chile
         \and
      	Instituto de Alta Investigaci\'on, Universidad de Tarapac\'a, Casilla 7D, Arica, Chile
      	 \and
        Research School of Astronomy and Astrophysics, Australian National University, Canberra, ACT 2611, Australia
          \and
        Astronomisches Rechen-Institut, Zentrum f\"{u}r Astronomie der Universit\"{a}t Heidelberg, M\"{o}nchhofstr. 12-14, 69120 Heidelberg, Germany
   }
 
   \date{Received ...; accepted ...}
  
 
 \abstract%
{}
   {We aim to analyze the chemical evolution of the Small Magellanic Cloud adding 12 additional clusters to our existing sample having accurate and homogeneously derived metallicities. We are particularly interested in seeing if there is any correlation between age and metallicity for the different structural components to which the clusters belong, taking into account their positions relative to the different tidal structures present in the galaxy.}
   {Spectroscopic metallicities of red giant stars are derived from the measurement of the equivalent width of the near-IR calcium triplet lines. Cluster membership analysis was carried out using criteria that include radial velocities, metallicities, proper motions and distance from the cluster center.  }
   {The mean cluster radial velocity and metallicity were determined with a typical error of 2.1 km s$^{-1}$ and 0.03 dex, respectively. We added this information to that available in the literature for other clusters studied with the same method, compiling a final sample of 48 clusters with metallicities homogeneously determined. Clusters of the final sample are distributed in an area of $\sim$ 70 deg$^2$ and cover an age range from 0.4 Gyr to 10.5 Gyr. This is the largest sample of spectroscopically analyzed SMC clusters available to date.}
   {We confirm the large cluster metallicity dispersion ($\sim$ 0.6 dex) at any given age in the inner region of the SMC. The metallicity distribution of our new cluster sample shows a lower probability of being bimodal than suggested in previous studies. The separate chemical analysis of clusters in the six components (Main Body, Counter-Bridge, West Halo, Wing/Bridge, Northern Bridge and Southern Bridge) shows that only clusters  belonging  to  the  Northern  Bridge  appear to trace a V-Shape, showing a clear inversion of the metallicity gradient in the outer regions. There is a suggestion of a metallicity gradient in the West Halo, similar to that previously found for field stars. It presents, however, a very large uncertainty. Also, clusters belonging to the West Halo, Wing/Bridge and Southern Bridge exhibit a well-defined age-metallicity relation with  relatively  little  scatter  in  abundance  at  fixed  age compared  to  other  regions.}


\maketitle
\section{Introduction}
\label{sec:intr}

The Magellanic Clouds (MCs) constitute the pair of interacting galaxies closest to the Milky Way (MW). They consist of the Large Magellanic Cloud (LMC) and the Small Magellanic Cloud (SMC), located at distances of $\sim$ 49.59 $\pm$ 0.09 kpc \citep{pietrzynski+19} and 62.44 $\pm$ 0.47 kpc \citep{Graczyk+20} from the MW, respectively.  The MCs are  embedded within a diffuse structure of HI gas, where different components can be identified, such as the Magellanic Stream and the Leading Arm \citep{Mathewson+74,Putman+03,nidever+08,nidever+10,donghia+16}. These features have been interpreted as a consequence of the LMC-SMC interactions and/or the interaction of the two galaxies with the MW \citep{besla+10,besla+12,diaz+11,diaz+12}.  For a long time it was thought that  the MCs had orbited the MW multiple times \citep{kallivayalil+06a,kallivayalil+06b}. However, more recent work strongly suggests that both galaxies are experiencing their first close encounter with our Galaxy \citep{besla+07,piatek+08,kallivayalil+13,gaia2016,patel+17}, based on the latest accurate measurements of proper motions with the Hubble Space Telescope (HST) and Gaia. 

Several authors found evidence that the stellar populations of the SMC have been subjected to substantial perturbations by forces associated with the LMC (e.g.,  \citealt{evans+08,haschke+12a,dobbie+14a,subramanian+17,deleo+20}) and reveal complex patterns of velocities consistent with the idea of the SMC being in the process of tidal disruption (e.g.,  \citealt{niederhofer+18,Niederhofer+21,zivick+18}). Also stellar tidal tails have been found around both MCs \citep{besla+16,belokurov+17,pieres+17,mackey+18,belokurov+19,nidever+19,gaia2021,ElYoussoufi+2021}. \\

Close encounters between gas-rich galaxies will produce enhancement of star formation (e.g., \citealt{whitmore+99}) 
and subsequent chemical enrichment (e.g., \citealt{dacosta91,dopita+97,pagel+98}). These processes alter, for example, the age and spatial distributions of the stellar populations \citep{glatt+10,nayak+16,nayak+18,bitsakis+18,rubele+18}, as well as the metallicity distribution and its gradient \citep{Cioni09}. In particular, if we want to understand not only the chemical evolution but also other processes like star formation and kinematics in the MCs, it is necessary to have a description of the global dynamics of the Magellanic System, and vice versa: these processes give important information on parameters related to interactions.\\

Our group is carrying out a long term investigation of SMC clusters and field stars using CaII triplet (CaT) spectroscopy  (\citealt{parisi+09,parisi+10,parisi+15,parisi+16}, hereafter P09, P10, P15 and P16, respectively). As shown by \citet[][hereafter C04]{cole+04}, the CaT is a very efficient and accurate metallicity indicator, with minimal age effects \citep{carrera+13} and independent of the chemical evolution histories \citep{dacosta16}. 

These investigations led to several intriguing results, indicating some surprising differences between the clusters and field star population and raising a number of important questions, including: how real is the cluster metallicity spread at a given age? Does it vary with age? If real, is it a global effect or does it vary with radius/position in the galaxy? Is the cluster metallicity distribution (MD) indeed bimodal? Why do clusters and field stars present different MDs? Is there a cluster metallicity gradient (MG) or not? If so, does it really invert in the outer regions? This latter question in particular is still a very controversial issue. While SMC field stars show an unquestionable MG (P16, \citealt{dobbie+14a,choudhury+18,choudhury+20}), for the case of the star clusters the MG is not statistically significant (P15).  At the same time, \citet{parisi+14} did not find a clear age gradient (AG) from a sample of 50 SMC clusters.\\

On the other hand, \citet[][hereafter D14 and D16b, respectively]{dias+14,dias+16b} introduced the idea that both the AG and the MG, as well as the large dispersion of metallicities, clearly evident in our AMR (P15), are due to the fact of analyzing the complete sample of clusters, without taking into account their membership positions in different components of the galaxy with potentially different chemodynamic histories. Individual clusters should be studied as part of the respective sky region as these regions may have been created during the perturbed evolution of the SMC. Using the projected distance $a$ \citep{piatti+05}, D16b suggest that cluster samples should be divided taking into account the tidal morphological characteristics of the SMC. Specifically, D16b divide the catalog of \citet{Bica+08} in four groups  depending on whether their positions in the galaxy match the SMC main body ($a$ < 2$^\circ$), the region in which the Wing/Bridge is located, the Counter-Bridge or the West Halo. These last three regions are located in the outer part of the galaxy ($a$ > 2$^\circ$). They have a clear gas counterpart \citep{besla11} and  have been predicted by different models and simulations, as described above.

Since then, more details have been added to this framework. For example, \citet{belokurov+17} showed that the stellar counterpart of the Magellanic Bridge \citep{irwin+85,omkumar+21,dias+21}, widely known to be related to the gaseous bridge containing predominantly younger stars, has a separate Southern branch traced by RR Lyrae stars, i.e., older stellar populations also connecting the SMC to the LMC, although the reality of this feature is in dispute \citep{jacyszyn+20}. \citet{dias+21} (hereafter D21) using full phase-space information, revealed that the Magellanic Bridge has a third, Northern branch, with clusters moving towards the LMC, which confirms previous indications by, e.g., \citet{nidever+17}. D21 also found the first confirmed star cluster belonging to the tidal counterpart of the Magellanic Bridge, the so-called Counter-bridge \citep{diaz+12,muller+07,ripepi+17,muraveva+18,omkumar+21,dias+21,Niederhofer+21}. Finally, D16b have defined the West Halo, a separate structure that seems to be moving away from the SMC as well. This outward motion was confirmed by proper motions \citep{zivick+18,niederhofer+18}, and \citet{tatton+21} discussed the possibility that the West Halo is actually the beginning of the Counter-Bridge that warps behind the SMC towards the Northeast.\\

We believe that, in order to better help constrain answers to the previously raised questions, it is necessary not only to significantly increase the sample of clusters homogeneously studied, but also to analyze the chemical properties of the SMC star cluster system in the context of its dynamical history, i.e.,  to study the clusters recognizing the different present-day environments rather than treating
them all together. With this goal in mind, we here add a sample of 12 massive clusters, which belong to several different components, studied in exactly the same way as our previous clusters. \\

In Section \ref{sec:obs} we describe the observations and the reduction process carried out for the new cluster observations. The measurement of radial velocities and equivalent widths is described in Section \ref{sec:rv_ew}. Section \ref{sec:met_det} is dedicated to the detailed description of the metallicity determinations. The analysis of the MG and AMR in the SMC is carried out in Section \ref{sec:met_analysis}. Finally, in Section \ref{sec:summary} we summarize our conclusions.

\section{Spectroscopic observations and reduction}
\label{sec:obs}

\begin{figure}[!htb]
    \centering
    \includegraphics[width=\columnwidth]{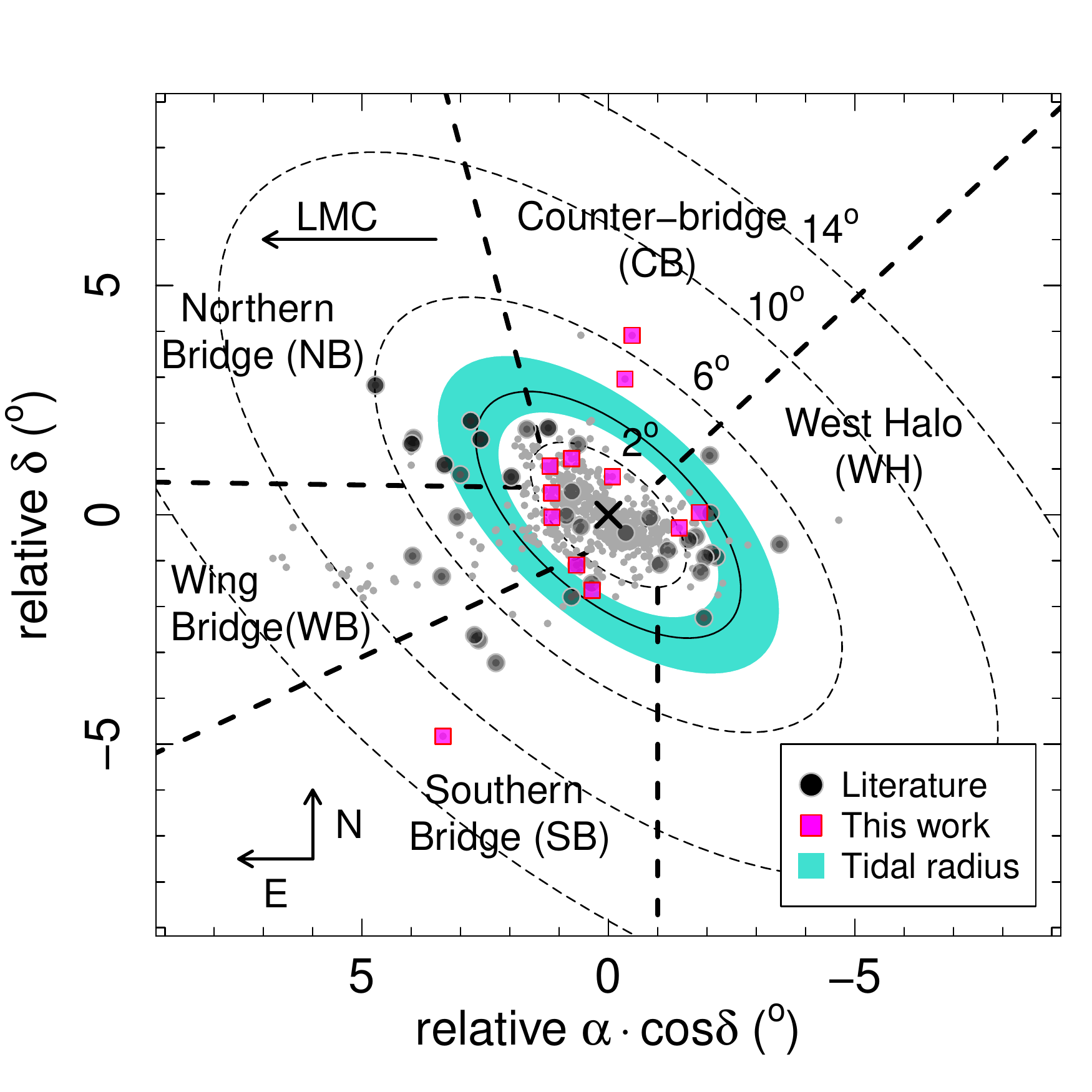}
    \caption{Projected distribution of SMC star clusters from the catalog of \citet{bica+20} represented by grey dots. Black circles are clusters previously studied with CaT  and pink squares  clusters studied in this work. Thin dashed lines indicate the ellipses used as a proxy for the distance to the SMC center. The distance $a$ is the semi-major axis of the ellipses indicated in degrees in the figure. The ellipses are tilted by $45^{\circ}$ and have an aspect ratio of $b/a=0.5$. Thick dashed lines split the regions outside $a>2^{\circ}$  into different SMC components. The SMC tidal radius of $a=3.4^{\circ}$ $^{+1.0}_{-0.6}$ (D21) is shown in turquoise. 
    }
    \label{fig:2Dsky}
\end{figure}

The observational data used in the present work were downloaded from the ESO Archive\footnote{http://archive.eso.org/cms.html}. The observations were carried out in 2005 and 2006 at the  VLT at European Southern Observatory (ESO, Paranal, Chile), under the programs 075.B-0548 and 073.B-0488 in service mode (PI: Eva Grebel). The SMC is the only nearby galaxy that has formed and preserved populous star clusters seemingly continuously over the past $\sim$ 11 Gyr. Therefore, for this program, populous SMC clusters were selected with prominent red giant branches that sampled and covered the past 11 Gyr, helping to provide a well-sampled AMR and quantifying metallicity spreads at any given age among  intermediate-age clusters in the galaxy. Preliminary results of the analysis of these data were presented by \citet{kayser+07}. The selected clusters are listed in Table \ref{tab:sample}, where we show their different identifications as well as their corresponding equatorial coordinates, the adopted cluster age, mass and the semi-major axis $a$. We adopted the designations of the different SMC components of D16b updated by D21 (shown in Fig. \ref{fig:2Dsky}) and associate our cluster sample with these components based on the projected line-of-sight locations of our clusters.  The resulting associations with the components are listed in the last columns of Table \ref{tab:sample}

\begin{table*}[!htb]
\caption{SMC cluster sample}             
\label{tab:sample}      
\centering                          
\footnotesize
\begin{tabular}{l c c c c c c c}        
\hline\hline                 
\noalign{\smallskip}  
  \multicolumn{1}{l}{Cluster} &
  \multicolumn{1}{c}{RA (J2000.0)} &
  \multicolumn{1}{c}{DEC (J2000.0)} &
  \multicolumn{1}{c}{Age} &
  \multicolumn{1}{c}{$\log(M/M_{\odot})$} & 
  \multicolumn{1}{c}{Ref.} &
  \multicolumn{1}{c}{a$^*$} &
  \multicolumn{1}{c}{component} \\
    \multicolumn{1}{c}{} &
  \multicolumn{1}{c}{($h$ $m$ $s$)} &
  \multicolumn{1}{c}{($^{\circ}$ ' '')} &
  \multicolumn{1}{c}{(Gyr)} &
  \multicolumn{1}{c}{} &
  \multicolumn{1}{c}{} &
  \multicolumn{1}{c}{($^{\circ}$)} &
  \multicolumn{1}{c}{} \\
\noalign{\smallskip}
\hline 
\noalign{\smallskip}
L43, K28, ESO 51-4                                          & 00:51:39.55  & -71:59:56.6  & 2.1 $\pm$ 0.5   &$4.7_{-0.41}^{+0.23}$&  1,8   & 1.4   &  MB\\ \\
K44, L68, RZ 135                                            & 01:02:06.34  & -73:55:22.7  & 2.0 $\pm$ 0.3	&$5.09_{-0.47}^{+0.29}$  & 8,12	& 2.5 & SB\\ \\
L11, K7, ESO 28-22                                          & 00:27:45.17  & -72:46:52.5  & 3.0 $\pm$ 0.4   &$4.20 \pm 0.16$&  2,9   & 3.0   &  WH \\ \\
L32, ESO 51-2                                               & 00:47:24.00  & -68:55:12.0  & 4.8 $\pm$ 0.5   &$3.57 \pm 0.13$&  1,9   & 6.7   &  CB\\ \\
L38, ESO 51-3, OGLE 308                                               & 00:48:50.00  & -69:52:12.0  & 6.5 $\pm$ 0.5   &$4.70$&  3,11   & 5.0   &  CB \\ \\
L116, ESO 13-25, AM 0155-775,                                & 01:55:33.00  & -77:39:18.0  & 2.8 $\pm$ 1.0   & -- &  1,9   & 11.6  &  SB   \\
 OGLE 91                                             &              &                               &                      &        &       &    \\ \\
NGC 152, L15, K10, ESO 28-24                                & 00:32:56.26  & -73:06:56.6  & 1.27 $\pm$ 0.07 &$4.80_{-0.41}^{+0.23}$&  4,8   & 2.0   &  WH  \\ \\
NGC 339, L59, K36, ESO 29-25                                & 00:57:48.90  & -74:28:00.2  & 6.5 $\pm$ 0.5   &$4.76_{-0.23}^{+0.15}$/$5.68$&  5,10,11   & 2.9   &  SB \\ \\
NGC 361, L67, K46, ESO 51-12                                & 01:02:12.83  & -71:36:16.2  & 6.5 $\pm$ 0.5   &$4.49_{-0.44}^{+0.25}$&  6,8    & 1.5   &  MB \\ \\
NGC 411, L82, K60, ESO 51-19,                               & 01:07:55.95  & -71:46:04.5  & 1.38            &$4.48_{-0.24}^{+0.12}$&  7,10   & 1.6   &  MB \\
RZ 172                                              &              &                               &                      &        &       &    \\ \\
NGC 416, L83, K59, ESO 2932,                                & 01 07 59.00  & -72:21:20.0 & 6.0 $\pm$ 0.5   &$4.81_{-0.44}^{+0.26}$/$4.90_{-0.04}^{+0.11}$/$5.53$&  5,8,10,11   & 2.6   &  CB\\ 
OGLE-CL SMC 158                                             &              &                               &                      &        &       &    \\ \\
NGC 419, L85, K58, ESO 29-33,                               & 01:08:17.79   & -72:53:02.8  & 1.6 $\pm$ 1.0   &$5.16_{-0.40}^{+0.22}$/$4.80_{-0.12}^{+0.09}$&  3,8,10   & 1.9   &   MB \\
LI-SMC 182, OGLE-CL SMC                                     &              &                               &                      &        &       &    \\ \\

\noalign{\smallskip}
\hline                                   
\end{tabular}\\
\textbf{References:} (1) \citet{piatti+01}, (2) \citet{livanou+13}, (3) \citet{glatt+08b}, (4) D16b, (5) \citet{Lagioia+19}, (6) \citet{mighell+98}, (7) \citet{li+16}, (8) \citet{gatto2021}, (9) \citet{santos+20}, (10) \citet{song+21}, (11) \citet{glat2011}, (12) \citet{parisi+14}. * The projected distance $a$ is the semi-major axis of the ellipse of axis ratio $b/a = 0.5$ coincident with the cluster position \citep{piatti+05}.

\end{table*}

\begin{figure}[!htb]
    \centering
    \includegraphics[width=\columnwidth]{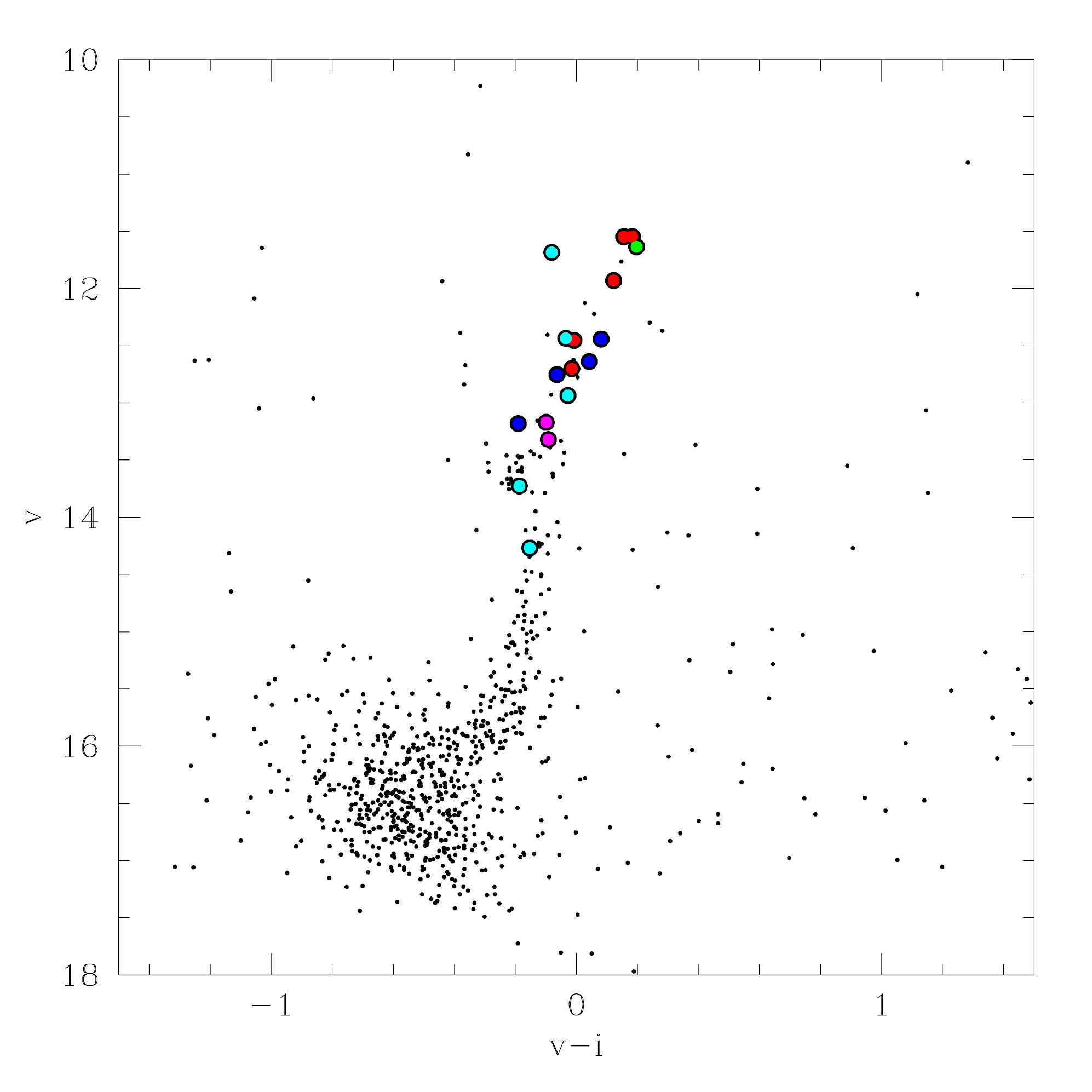}
    \caption{Instrumental color-magnitude diagram of the cluster L\,38. The spectroscopic targets are marked with large circles. Blue, cyan, green and magenta symbols represent stars discarded as cluster members because of their distance from the cluster center, RV, metallicity and PM values. Red circles show the adopted cluster members. See Section \ref{sec:members} for details. 
    }
    \label{fig:l38_CMD}
\end{figure}

\begin{figure}[!htb]
    \centering
    \includegraphics[width=\columnwidth]{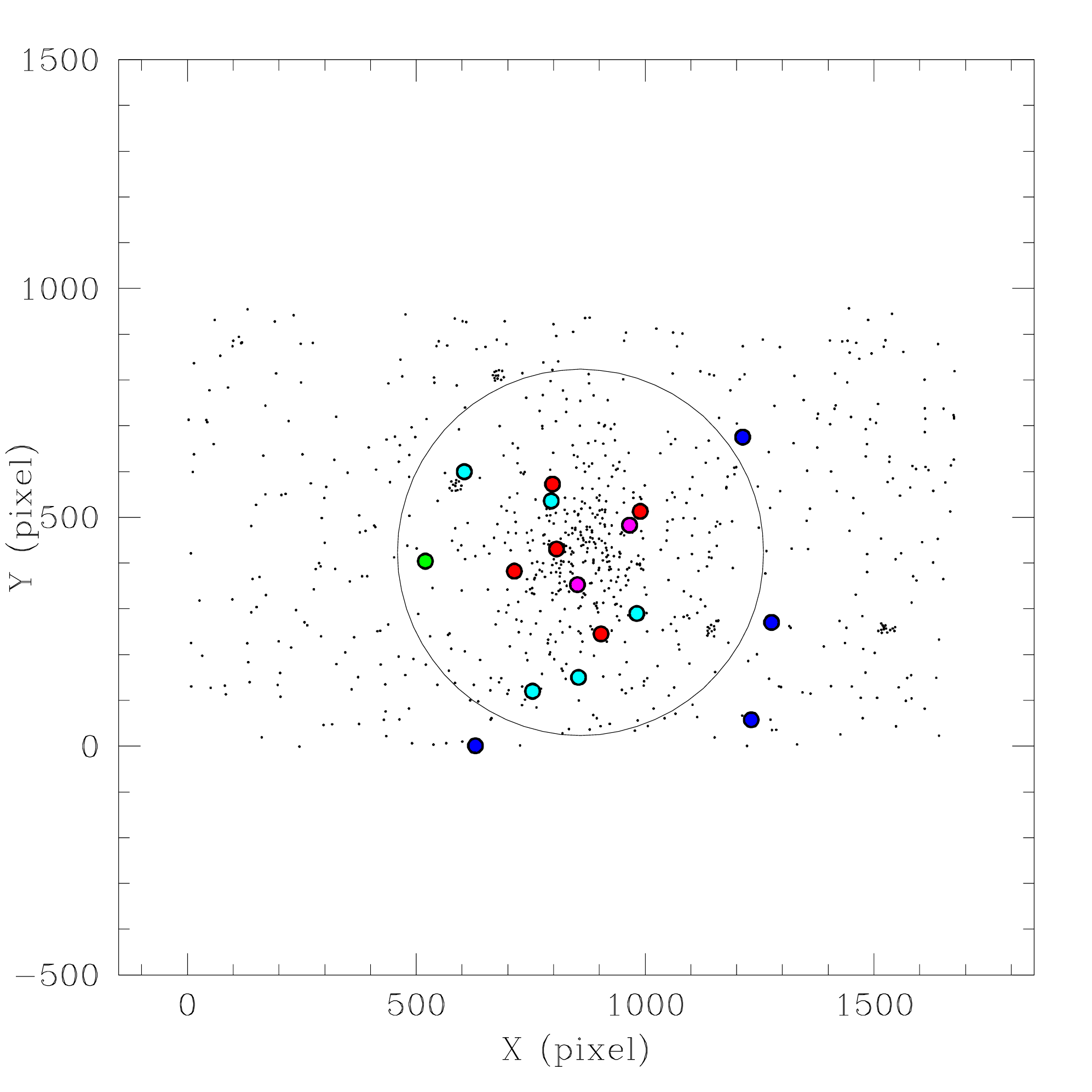}
    \caption{Chart of cluster L\,38. Spectroscopic targets are plotted with the same color code as in Fig. \ref{fig:l38_CMD}. The adopted cluster radius is represented by the circle. 
    }
    \label{fig:l38_chart}
\end{figure}

In each cluster, spectroscopic targets correspond to red giant stars belonging to the clusters and their surrounding fields. As example, for the cluster L\,38, we show in Figs. \ref{fig:l38_CMD} and \ref{fig:l38_chart} the location of selected targets in the cluster color-magnitude diagram (CMD) and positional chart. The positions and magnitudes needed to build these figures were obtained from PSF photometry performed on the $V$ and $I$ pre-images taken previously with the same instrument. 

Using the FORS2 instrument on the VLT, spectra of 502 purported cluster red giant stars were obtained. The instrument, in mask exchange units (MXU) mode, was used with the 1028z+29 grism and OG590+32 filter. FORS2 has two CCDs, the master and the secondary chips, of a size of 2000 $\times$ 4000 pixels each. In all cases the master CCD was centered on the cluster and the slave will therefore contain a much higher fraction of field stars.
Between 19 and 36 slits (1'' wide) were located in the total frame. Pixels were binned 2$\times$2, yielding a plate scale of 0.25'' pixel$^{-1}$ and a dispersion of $\sim$ 0.85 \AA \space pixel$^{-1}$.  The spectral range covered by the resulting spectra is  1750 \AA~ (7750 $-$ 9500 \AA), with a central wavelength coincident with the region of the CaT lines ($\sim$ 8600 \AA). Observations were made with 475 sec of exposure time.

 We performed the bias, flat-field, distortion correction and the wavelength calibration using the pipeline provided by ESO (version 2.8). The necessary calibration images were acquired by the ESO staff. The mentioned pipeline also performed the extraction and the sky subtraction. The IRAF tasks {\it scombine} and {\it continuum} were used for the combination of the spectra and the normalization of the combined spectra, respectively. We note that both the instrument as well as reduction procedure are the same as used in our previous work.

\section{Radial velocity and equivalent width measurements}
\label{sec:rv_ew}

We have followed the prescriptions detailed in our previous work (P09, P15) to perform the measurement of the radial velocities (RVs) of our targets. The IRAF {\it fxcor} task was used to calculate the cross-correlation between the observed stars and RV templates. The selected template stars are taken from C04, and are the same red giant spectra used in our previous work (see, for example, \citealt{grocholski+06} and P09 for more details). 

In order to also maintain consistency with our previous work, equivalent widths (EWs) were measured on the normalized spectra by fitting a combination of a Gaussian and a Lorentzian function. The spectra were previously corrected for the Doppler effect using  our measured values of the observed RVs. EWs were determined considering the line and continuum bandpasses from \citet{armandroff+88}. 

\section{Metallicity determination}
\label{sec:met_det}

\subsection{Calcium Triplet Calibrations}
\label{sec:cals}

In the literature there is a vast amount of work that clearly establishes the correlation between the sum of EWs of the CaT lines ($\Sigma$EW),  $v-v_{HB}$  and metallicity (\citealt{dias+20} and references therein). Several authors have proposed the use of the so-called reduced equivalent width ($W'$), which removes the dependence of the $\Sigma$EW on the effective temperature and surface gravity \citep{armandroff+91,olszewski+91}.  This correction requires the use of the differential magnitude (in a given photometric system) between the observed star and the horizontal branch (or clump). As a consequence, in the luminosity-$\Sigma$EW plane, stars of the same cluster should fall along a straight line, having a slope of the same value for all clusters, but the lines will  be displaced vertically in the mentioned plane according to the cluster metallicity. 
$W'$ has been calibrated by several authors considering different samples of calibration objects, not only for visual (\citealt{rutledge+97a}; C04; \citealt{carrera+13,saviane+12,dacosta16,dias+16a,vasquez+18}) but also for infrared magnitudes \citep{carrera+13,mauro+14,vasquez+15}, HST filters \citep{husser+20}, and the Gaia G-band \citep{simpson20}.  More generally, \citet[][hereafter DP20]{dias+20} analyzed the dependence of the CaT calibration for a wide variety of filters covering a wavelength range from 445 to 2135 nm ($BgVGriIzYJKs$, see their Table 4) and concluded that the calibration does not depend on the transmission of each filter but rather on their effective wavelength. They derived a generic function for all wavelengths in this range, highlighting  that redder filters constrain the calibration better.

For visual calibrations, in all cases except \citet{carrera+13}, $W'$ is defined as follows:
\begin{equation}
    \begin{aligned}
    W' = \Sigma EW + \beta_V \times (V-V_{HB}),
       \end{aligned}
    \label{eq:red_ew}
\end{equation}

\noindent forming the CaT index by the contribution of the three CaT lines ($\Sigma$EW $=$ EW$_{8489}$ $+$ EW$_{8542}$  $+$ EW$_{8662}$, C04), the two most intense lines ($\Sigma$EW $=$ EW$_{8542}$  $+$ EW$_{8662}$, \citealt{saviane+12,dacosta16,vasquez+18}) or the sum of the EW of the three CaT lines weighted by the errors \citep{rutledge+97a}. 
%

 
Traditionally, in our series of papers on CaT metallicities of SMC clusters and field stars (P09, P10, P15, P16), we have used  the calibration of C04. Using abundances for red giants in five globular clusters (on the \citealt{carretta+97} scale) and in six open clusters (on the \citealt{friel02} scale), C04 derived a linear correspondence between $W'$ and [Fe/H], with a rms scatter of 0.07 dex:
\begin{equation}
    \begin{aligned}
    {\rm [Fe/H]}_{C04} = -2.966 (\pm 0.032) + 0.362 (\pm 0.014) W'\\
       \end{aligned}
    \label{eq:met_c04}
\end{equation}

The $\beta_V$ value derived by C04 is 0.73 $\pm$ 0.04  that is valid in the ranges of $-$ 2$\leq$ [Fe/H] $\leq$ $-$0.2  and 2.5  $\leq$ (age/Gyr)  $\leq$ 13. Note that some of our clusters are younger than the minimum age limit for which the calibration is defined. However, \citet{carrera+07} showed that the influence of age is small, even for ages $<$ 1 Gyr.

%


%


It is generally accepted that the correspondence between  [Fe/H] and  $W'$ follows a linear behavior (e.g., C04; \citealt{dacosta16}), although certain indications have been found that this relation deviates from linearity for metallicities larger than -0.7 dex (\citealt{saviane+12,mauro+14,vasquez+15,dias+16a,vasquez+18}). In the present work, we also explore this possible effect. 

We  examine  the calibration of \citet[hereafter V18][]{vasquez+18} in order to compare  results, especially for clusters having $W'$ larger than $\sim$ 5 , where the calibration  apparently   becomes non-linear. For completeness, we also derive metallicities following  \citet[][hereafter DC16]{dacosta16}. 
%


\subsection{Metallicity and cluster membership determination}
\label{sec:members}

\begin{figure}[!htb]
    \centering
    \includegraphics[width=\columnwidth]{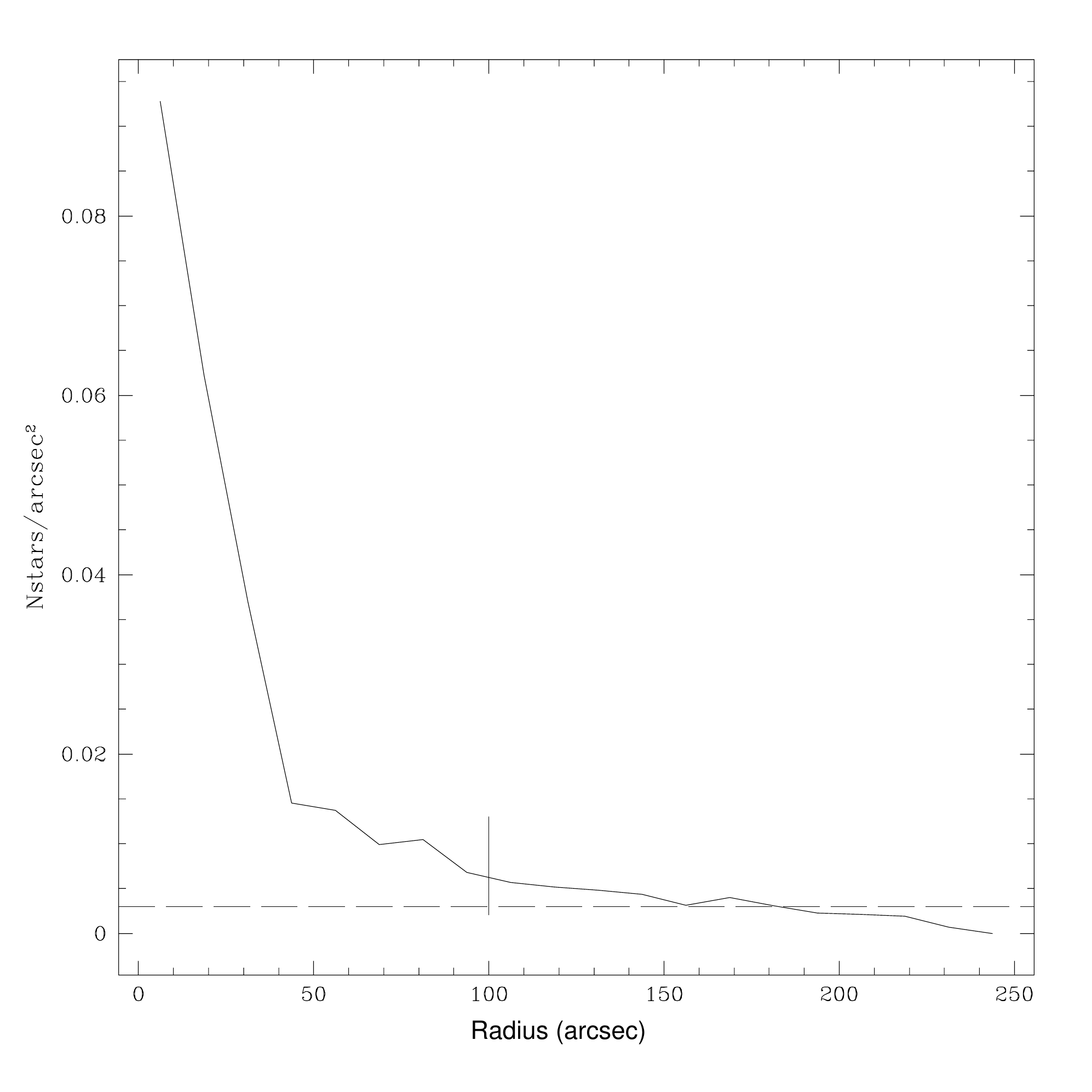}
    \caption{Radial stellar density profile of cluster L38. The x-axis represents the distance to the cluster center and the y-axis the projected stellar number density. The vertical line marks the adopted cluster radius and the horizontal dashed line is the background density.
    }
    \label{fig:l38_prf}
\end{figure}

As a first step, we calculated metallicities according to equation \ref{eq:met_c04}. For each target, the CaT index $\Sigma$EW was built by adding the EWs of the three CaT lines, in the same way as C04. We then calculated the $W'$ using C04's $\beta_v$ value of  0.73. For each cluster of our sample, $v$ magnitudes were obtained from PSF photometry performed  on the $V$ and $I$ pre-images. We then 
obtained the apparent cluster radius from the radial stellar density profile (based on star counts over the entire frame) and chosen as the distance from the cluster center where the stellar background density intersects the cluster density profile (see P09 and P15 for more details). We then adopted a conservative approach for cluster membership determination by adopting a cut off radius for membership 
that was approximately 2/3 of the apparent cluster radius. We adopted a smaller radius in order to maximize the cluster member probability with the goal to find the mean cluster metallicity. 
In Fig. \ref{fig:l38_prf}, we show the stellar radial density profile for the cluster L38. \\

($v$, $v-i$) CMDs of stars located within the apparent cluster radius  were constructed and used to derive the $v_{HB}$. We use lower-case letters to denote  that our photometry is uncalibrated. 
To determine $v_{HB}$, we located a box $v$ $\times$ $v-i$  of  0.7 $\times$ 0.3 mag centered on the red clump (RC) by eye and calculated $v_{HB}$ as the median value of stars located in the box. The error of $v_{HB}$ is the standard error of the median. We follow the same procedure for all clusters of our sample.  

\begin{figure}[!htb]
    \centering
    \includegraphics[width=\columnwidth]{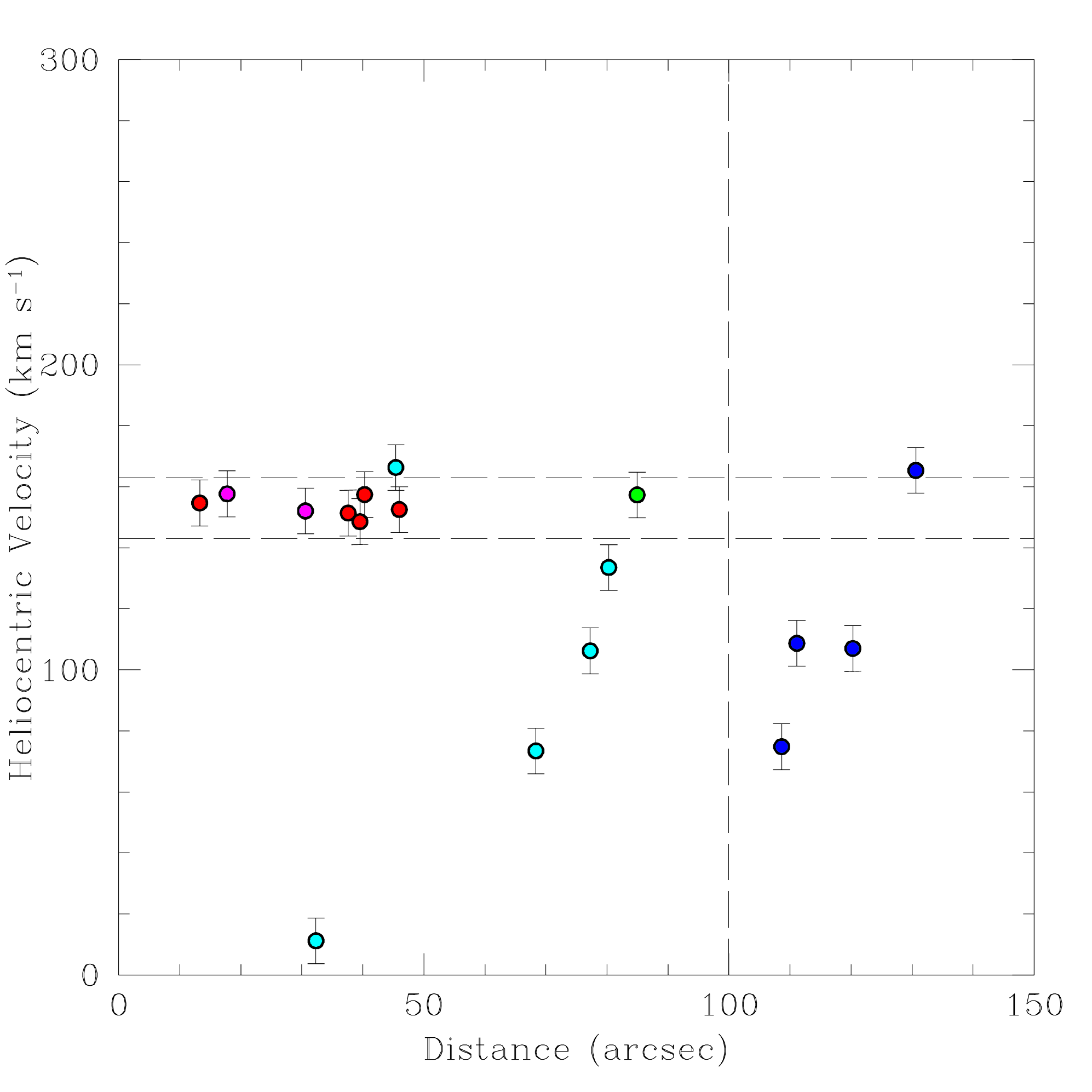}
    \caption{Radial velocity vs. distance from the cluster center for L38 targets. Radial velocity cuts and the adopted cluster radius are marked with horizontal and vertical lines, respectively. The color code is the same as in Fig. \ref{fig:l38_CMD}.
    }
    \label{fig:l38_rv}
\end{figure}

\begin{figure}[!htb]
    \centering
    \includegraphics[width=\columnwidth]{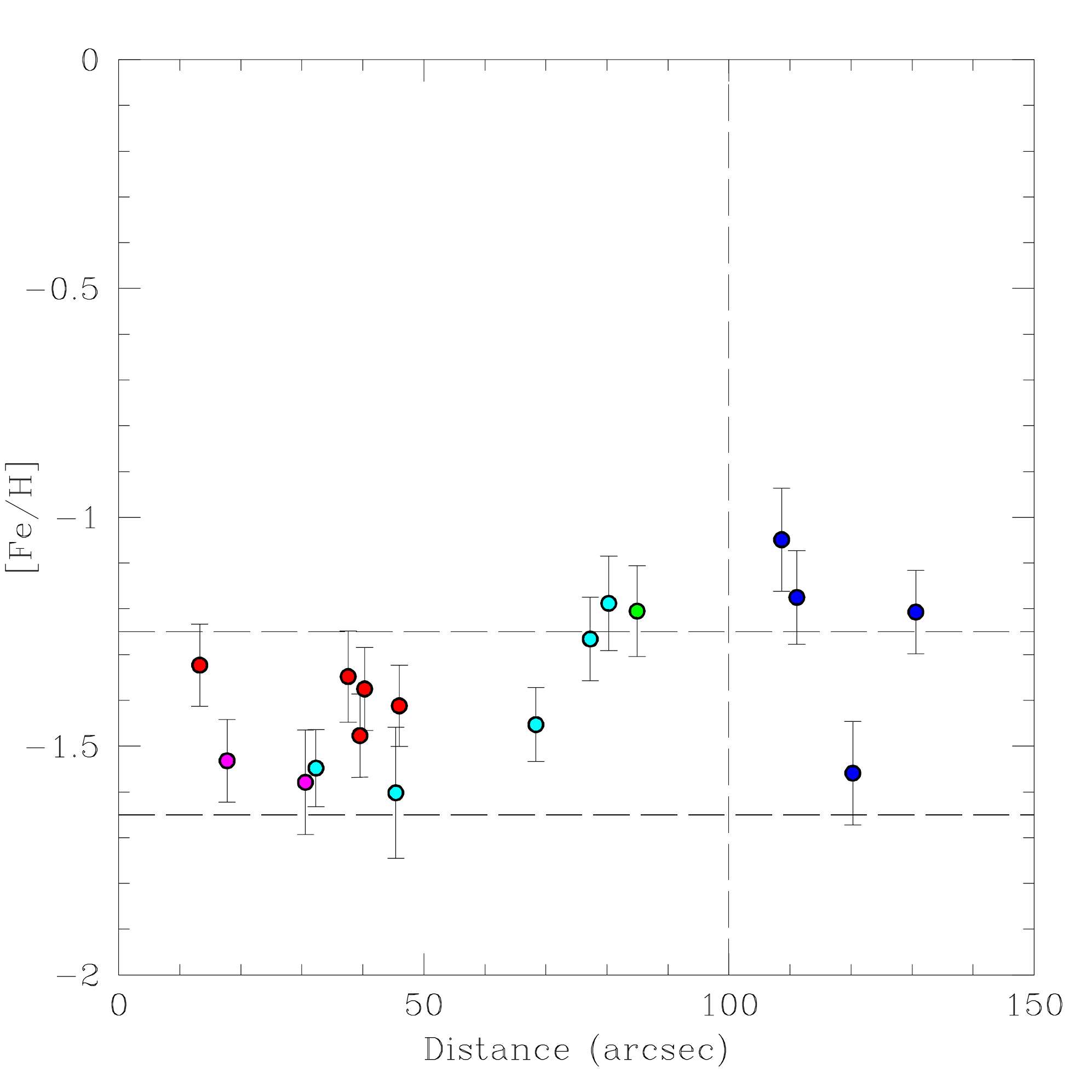}
    \caption{Metallicity vs. distance from the cluster center for L38 targets. Metallicity cuts and the adopted cluster radius are marked with horizontal and vertical lines, respectively. The color code is the same as in Fig. \ref{fig:l38_CMD}.
    }
    \label{fig:l38_met}
\end{figure}

For consistency with our previous work, we apply the cluster  membership method used by our group in P09 and P15. The distance of each star from the center of the cluster, the RV and the [Fe/H] are the parameters considered previously to establish whether the observed stars are likely members of the cluster or, conversely, belong to the  surrounding fields. Cluster members must be closer to the center than the adopted cluster radius. It is also expected that cluster stars have similar RV values within the RV dispersion expected for a cluster. Also, with a mean RV value being not necessarily similar to that of field stars and with a smaller dispersion. In addition, under the assumption that the clusters do not possess any intrinsic internal abundance dispersion, the observed dispersion in the 
cluster member metallicities should correspond to that expected from the individual metallicity errors. To illustrate our method, we show in Figs. \ref{fig:l38_rv} and \ref{fig:l38_met} the behavior  of the RV and metallicity  of the red giant stars observed in the cluster L38 vs. the distance from the cluster center, respectively.  

\begin{figure}[!htb]
    \centering
    \includegraphics[width=\columnwidth]{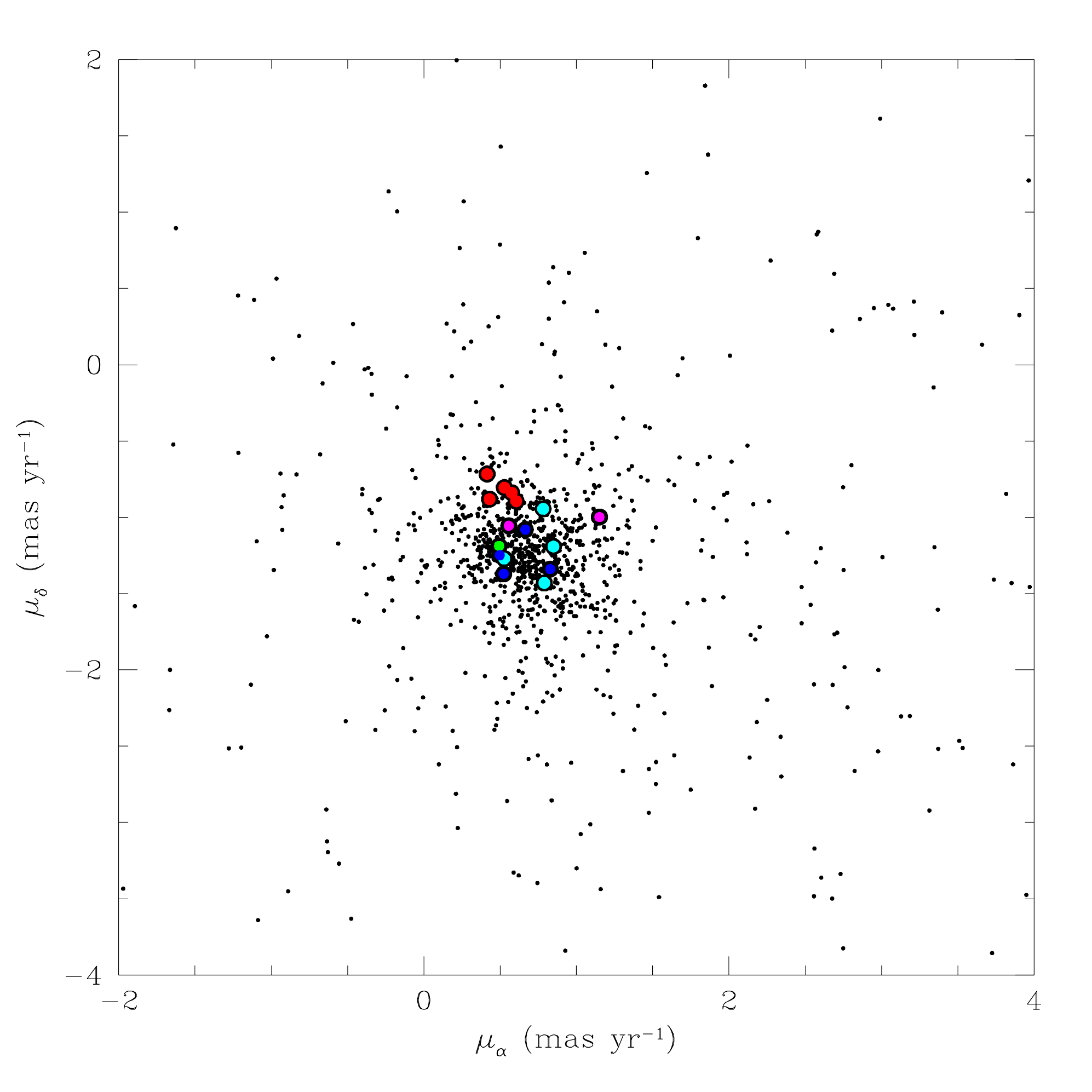}
    \caption{Proper motion plane for cluster L38. Black points represent stars from the Gaia eDR3 catalogue and large circles stand for  our spectroscopic targets.The color code is the same as in Fig. \ref{fig:l38_CMD}
    }
    \label{fig:l38_pm}
\end{figure}

In Figs. \ref{fig:l38_rv} and \ref{fig:l38_met}, the vertical line marks the adopted cluster radius and the horizontal lines are the corresponding cuts in RV and metallicity adopted in this work.  
The RV cuts are the sum in quadrature of the expected dispersion within a cluster ($\sim$ 5 km s$^{-1}$, \citealt{pryor+93}) and our error in the calculation of the RV 
values ($\sim$ 7.5 km s$^{-1}$) which amounts to 9 km s$^{-1}$. As we have done in all our previous work, we rounded up our RV cuts to $\pm$ 10 km s$^{-1}$. The metallicity cuts ($\pm$ 0.20 dex) correspond to the mean error of our metallicity determination for the observed red giants. We also limited adopted cluster members to have $v-v_{HB} < +$0.2  to avoid any possible effect on the metallicity due to the possible loss of linearity between $v-v_{HB}$ and $\Sigma$EW  below the magnitude of the RC.

 In order to improve our traditional membership analysis and select stars with the maximum probability of being cluster members, we also analyze the proper motions (PMs) of the observed stars. Using the corresponding cluster central coordinates, we searched for Gaia eDR3 \citep{gaia2021} stars in the  area of each cluster. We then identified our targets in the Gaia eDR3 astrometric catalog and we discarded those stars whose motion is not consistent with the average motion of the cluster in the PM plane [$\mu_{\alpha}$,$\mu_{\delta}$] (Fig. \ref{fig:l38_pm}). It is expected that cluster member stars have a similar movement within the uncertainties and 
the non-member stars present a greater dispersion in both coordinates. Therefore we discard those spectroscopic members with deviating PM with respect to the values that our 
spectroscopic members present. 

\begin{figure}[!htb]
    \centering
    \includegraphics[width=\columnwidth]{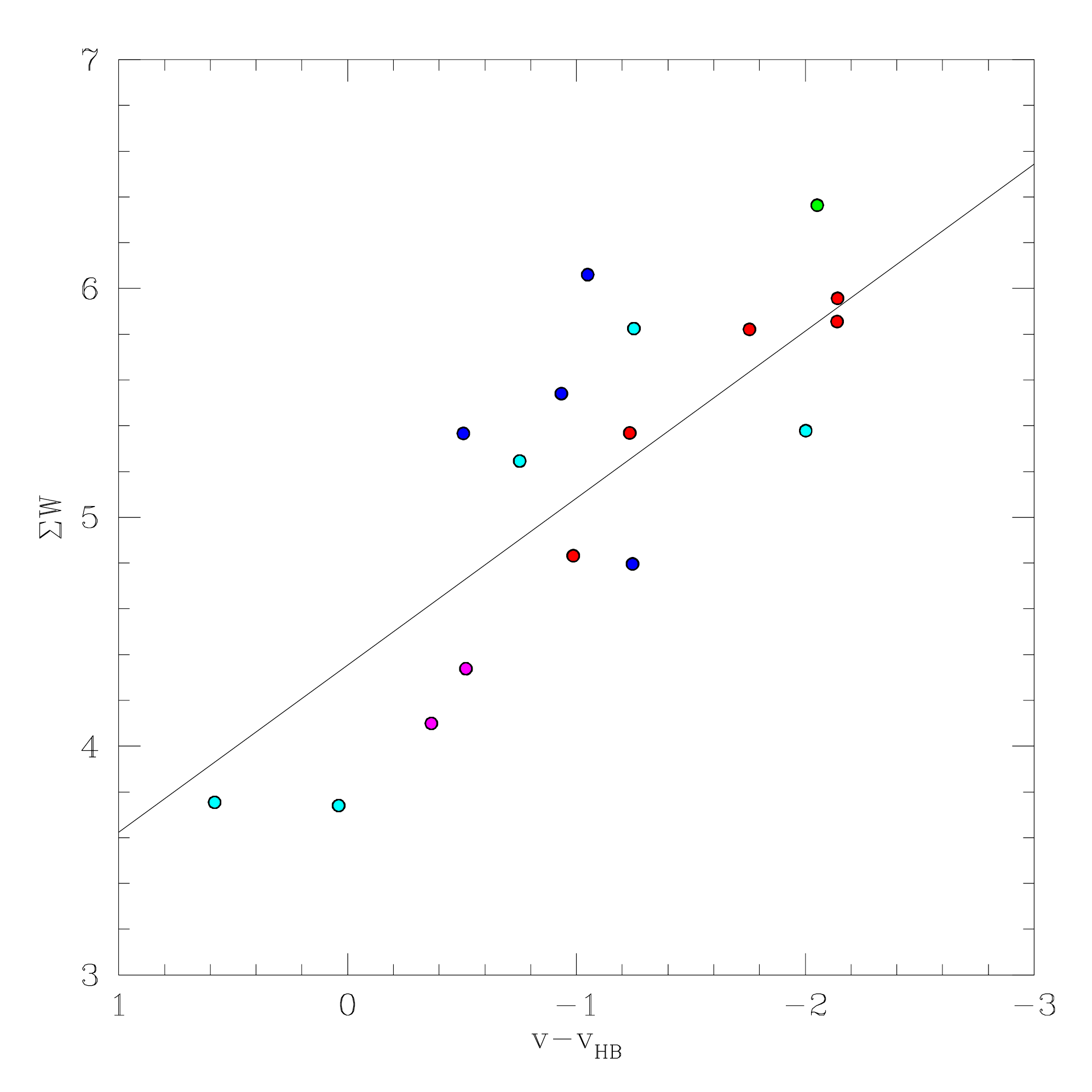}
    \caption{Sum of the equivalent widths of the three CaT lines vs. $v$-$v_{HB}$ for identified  members and non-members  of  cluster L38.  The  solid  line  represents the isometallicity line corresponding to the mean cluster metallicity. The color code is the same as in Fig. \ref{fig:l38_CMD}.
    }
    \label{fig:l38_slope}
\end{figure}

\begin{figure}[!htb]
    \centering
    \includegraphics[width=\columnwidth]{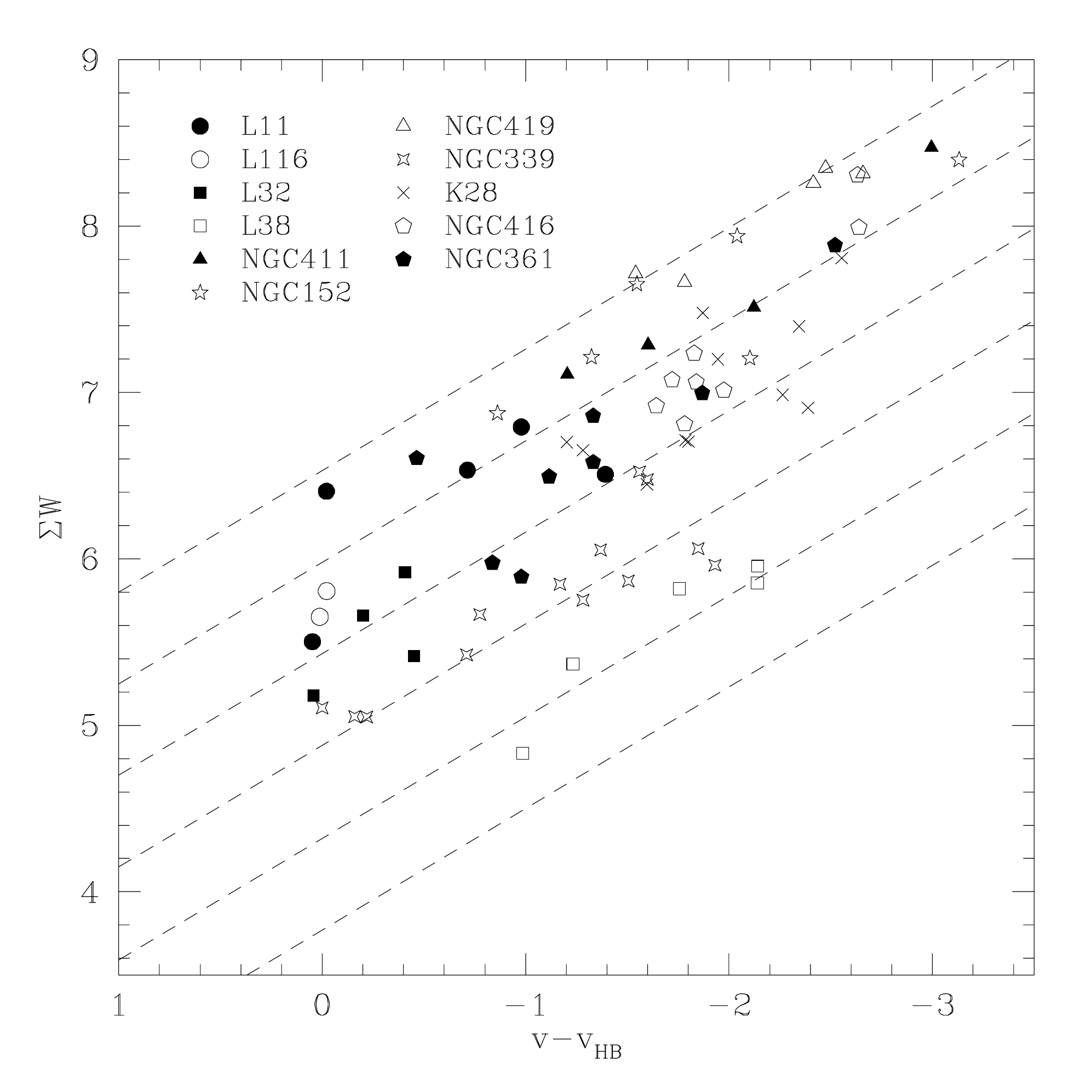}
    \caption{The same as in Fig. \ref{fig:l38_slope} but for members in all our cluster sample. The dashed lines represent lines of equal metallicity  of -0.6, -0.8, -1.0, -1.2, -1.4, -1.6 dex from top to bottom.
    }
    \label{fig:slope_all}
\end{figure}

The color code in the figures, which is explained in the caption of Fig. \ref{fig:l38_CMD}, is the same as in our previous work (\citealt{grocholski+06}, P09, P15). Observed stars that have passed the cuts in radius, RV, metallicity and PMs (red symbols) are considered our final cluster members.  Fig. \ref{fig:l38_slope} shows the CaT index $\Sigma$EW vs. $v-v_{HB}$ for stars observed in cluster L38, where the linear behavior of cluster member stars can be seen,  in contrast to the field stars whose distribution has a larger scatter.  In Fig. \ref{fig:slope_all}, we plot the complete sample, as in Fig. \ref{fig:l38_slope}, but including only our final members in each cluster.

In Table \ref{tab:individuals}, we list successively  for the member stars of our cluster sample: the identification of the star, equatorial coordinates, heliocentric RV,  $v-v_{HB}$, $\Sigma$EW and metallicity on the C04 scale, with their respective errors.

\begin{table*}[!htb]
\caption{Measured values for member stars}             
\label{tab:individuals}      
\centering                          
\footnotesize
\begin{tabular}{l l l l l l l l l l l l}        
\hline\hline                 
\noalign{\smallskip}  
  \multicolumn{1}{c}{Cluster member} &
  \multicolumn{1}{c}{RA (J2000.0)} &
  \multicolumn{1}{c}{DEC (J2000.0)} &
  \multicolumn{1}{c}{RV} &
   \multicolumn{1}{c}{$v-v_{HB}$} &
   \multicolumn{1}{c}{$\Sigma$EW} &
   \multicolumn{1}{c}{[Fe/H]$_{C04}$} &
   \multicolumn{1}{c}{$\mu_{\alpha}$} &
   \multicolumn{1}{c}{$\mu_{\delta}$}\\
     \multicolumn{1}{c}{} &
  \multicolumn{1}{c}{($h$ $m$ $s$)} &
  \multicolumn{1}{c}{($^{\circ}$ ' '')} &
  \multicolumn{1}{c}{(km s$^{-1}$)} &
   \multicolumn{1}{c}{(mag)} &
   \multicolumn{1}{c}{(\AA)} &
   \multicolumn{1}{c}{(dex)} & 
   \multicolumn{1}{c}{(mas yr$^{-1}$)} &
   \multicolumn{1}{c}{(mas yr$^{-1}$)}\\
\noalign{\smallskip}
\hline 
\noalign{\smallskip}
L38-6  & 0 48 48.89 & -69 52 58.80 & 152.6 $\pm$ 1.8 &	-2.14 &	5.86 $\pm$	0.12 &	-1.41 $\pm$	0.09 & 0.49 $\pm$ 0.16 & -0.84 $\pm$ 0.10\\
L38-11 & 0 48 58.12 & -69 52 26.40 & 151.4 $\pm$ 3.2 &	-1.23 &	5.37 $\pm$	0.18 &	-1.35 $\pm$	0.10 & 0.41 $\pm$ 0.27 & -0.99 $\pm$ 0.17\\
L38-13 & 0 48 53.63 & -69 52 12.00 & 154.7 $\pm$ 1.8 &	-1.76 &	5.82 $\pm$	0.12 &	-1.32 $\pm$	0.09 & 0.61 $\pm$ 0.18 & -0.83 $\pm$ 0.13\\
L38-15 & 0 48 44.70 & -69 51 50.40 & 148.6 $\pm$ 2.3 &	-0.99 &	4.83 $\pm$	0.16 &	-1.48 $\pm$	0.09 & 0.78 $\pm$ 0.28 & -0.77 $\pm$ 0.20\\
L38-17 & 0 48 54.05 & -69 51 36.00 & 157.4 $\pm$ 1.3 &	-2.14 &	5.96 $\pm$	0.12 &	-1.38 $\pm$	0.09 & 0.52 $\pm$ 0.13 & -0.90 $\pm$ 0.10\\

\hline                                   
\end{tabular}
\tablefoot{This table is available in its entirety in the online journal and in the CDS database. A portion is shown here for guidance regarding its form and content.
}
\end{table*}

Using only cluster member stars, we then calculated the mean cluster RV and [Fe/H]. In Table \ref{tab:results}, we present the cluster name in column (1) and the number $n$ of stars that turned 
out to be cluster members in column (2), as well as the mean cluster RVs and  metallicity, with their respective errors, in columns (3) and (4), respectively. We note that some clusters in our sample present smaller RV values (differences between $\sim$ 9 and $\sim$ 24 km s$^{-1}$) than those obtained by \citet{song+21}. A possible source of discrepancy may include the correction due to the offset of the stars in the slits, whose values are of the order of the observed differences.
Although these differences do not affect our membership selection or the metallicity analysis presented here, they must be be considered when using our RVs for any dynamical analysis.

\begin{table}[!htb]
\caption{Derived cluster mean properties}
\label{tab:results}      
\centering                          
\footnotesize
\begin{tabular}{l c c c}        
\hline\hline                 
Cluster  & n  & RV* & [Fe/H]$_{C04}$ \\    
 &  & (km s$^{-1}$)  & (dex)  \\
\hline                        
K28     & 11 & 132.80 $\pm$ 1.61 &  -0.94 $\pm$ 0.03  \\
K44     & 11 & 143.70 $\pm$ 0.83 &  -0.78 $\pm$ 0.03 \\
L11     &  5 & 126.28 $\pm$ 1.66 &  -0.83 $\pm$ 0.06  \\
L32     &  4 & 120.28 $\pm$ 3.37 &  -0.96 $\pm$ 0.04  \\
L38     &  5 & 152.93 $\pm$ 1.50 &  -1.39 $\pm$ 0.03  \\
L116    &  2 & 153.44 $\pm$ 2.55 &  -0.89 $\pm$ 0.02  \\
N152    &  6 &  148.41 $\pm$ 2.43 &  -0.72 $\pm$ 0.02  \\
NGC 339 & 13 & 103.30 $\pm$ 2.35 &  -1.15 $\pm$ 0.02  \\
NGC 361 & 11 & 161.18 $\pm$ 1.24 &  -0.90 $\pm$ 0.03   \\
NGC 411  &  4 & 140.72 $\pm$ 3.89 &  -0.74 $\pm$ 0.04  \\
NGC 416  &  8 & 138.04 $\pm$ 1.25 &  -0.85 $\pm$ 0.04  \\
NGC 419  &  5 & 171.48 $\pm$ 2.53 &  -0.62 $\pm$ 0.02  \\
\hline                                   
\end{tabular}
\tablefoot{* For the clusters of our sample common to \citet{song+21} our RV values are systematically smaller, probably due to the correction for off-centering of the stars in the slit. }
\end{table}



We summarize in Table \ref{tab:met_lit}  the most important previous metallicity determinations for our sample. 
In general terms, there is a reasonable agreement with the metallicity values previously determined by other authors.

\begin{table*}[!htb]
\caption{Metallicity from the literature for our observed clusters}             
\label{tab:met_lit}      
\centering                          
\footnotesize
\begin{tabular}{l l l l }        
\hline\hline                 
\noalign{\smallskip}  
  \multicolumn{1}{c}{Cluster} &
   \multicolumn{1}{c}{[Fe/H]} &
   \multicolumn{1}{c}{method} &   
   \multicolumn{1}{c}{Reference} \\
 \noalign{\smallskip}
\hline 
\noalign{\smallskip}
K28     & -1.2 $\pm$ 0.2  & Washington photometry &\citet{piatti+01}\\
        & -1.0 $\pm$ 0.02 & integrated spectra &\citet{piatti+05} \\ \\
K44     & -1.1 $\pm$ 0.2 & Washington photometry & \citet{piatti+01} \\
     & -0.81 $\pm$ 0.04 & CaII triplet & P15 \\ \\

L11     & -0.93            & photometry &\citet{mould+92}  \\
        & -0.81 $\pm$ 0.13 &  CaII triplet &DH98\\
        & -0.8/-1.3        & Strömgren photometry &\citet{livanou+13} \\ \\

L32     & -1.2 $\pm$ 0.02  & Washington photometry&\citet{piatti+01}\\ \\

L38     & -1.65 $\pm$ 0.2  & Washington photometry &\citet{piatti+01}\\ \\

L116    & -1.1 $\pm$ 0.2  & Washington photometry &\citet{piatti+01}\\ \\

N152    & -1.25 $\pm$ 0.25  & integrated photometry &\citet{bica+86}     \\
        & -1.1 $-$ -1.4 &  integrated spectra &\citet{dias+10}\\
        & -0.73 $\pm$ 0.11 & high-resolution spectroscopy &\citet{song+21} \\ \\

NGC 339 & -1.19 $\pm$ 0.12 & CaII triplet& DH98 \\
        & -0.70            & Spectral indices -integrated colours& \citet{deFreitas+98}\\
        & -1.50            & photometry & \citet{mighell+98} \\
        & -1.10 $\pm$ 0.03 & Strömgren photometry& \citet{narloch+21} \\
        & -1.01 0.17 $\pm$ & high-resolution spectroscopy& \citet{song+21} \\ \\

NGC 361 &-1.25 $\pm$ 0.2  & integrated photometry& \citet{bica+86}\\
        & -0.8/-0.7       & Strömgren photometry& \citet{livanou+13}\\
        & -1.45             & photometry& \citet{mighell+98}\\
        & -0.7 $-$ -1.0    &integrated spectra& \citet{dias+10}\\
        & -0.79 $\pm$ 0.04 & Strömgren photometry& \citet{narloch+21} \\
        &-0.75 $\pm$ 0.17 & Strömgren photometry& \citet{narloch+21} \\ \\

NGC 411  & -0.7 $\pm$ 0.2 & integrated spectra& \citet{piatti+05} \\
         & -0.66 $\pm$ 0.09 & integrated spectra& \citet{piatti+05} \\ \\

NGC 416   & -1.1             & spectroscopy& \citet{martocchia+20} \\
         & -1.0 $-$ -1.2    & integrated spectra& \citet{dias+10} \\
         & -0.80 $\pm$ 0.17 & integrated spectra& \citet{dias+10} \\ \\

NGC 419  & -0.7 $\pm$ 0.3 & photometry& \citet{duran+84} \\
         & -1.2           & integrated photometry& \citet{bica+86}\\
         & -0.6           & Spectral indices -integrated colours& \citet{deFreitas+98}\\
         & -0.6 $-$ -1.4    & integrated spectra& \citet{dias+10}\\
         & -0.70            & spectroscopy& \citet{martocchia+17}\\
         & -0.84 $\pm$ 0.19 & high-resolution spectroscopy& \citet{song+19}\\
         & -0.66 $\pm$ 0.15 & high-resolution spectroscopy&  \citet{song+19}\\
\noalign{\smallskip}
\hline                                   
\end{tabular}
\end{table*}

\section{Results and discussion}
\label{sec:met_analysis}

In order to analyze the chemical evolution of the SMC, with a statistical sample as large as possible and with homogeneously determined metallicities, we have compiled from the literature all 
the clusters that have metallicities determined by the CaT technique. Besides the 12 clusters studied in the present work, we included in the sample the ones  previously studied by 
\citet[hereafter DH98][]{dacosta+98}, P09, P15 and D21. We have then a sample of 48 clusters with metallicities determined in a homogeneous way,  which represents a 40\% increase over the sample analyzed in P15. This sample naturally also substantially improves our coverage of the different SMC components, helping to clarify any individual trends. We present in Table \ref{tab:cat_sample}  the adopted metallicity, age, semi-major axis and component values for the additional cluster sample. With this extended cluster sample, we analyze the MG, MD and AMR in the SMC.\\

Two of the clusters of our sample (L11 and NGC339) possess CaT metallicities from DH98. Both CaT determinations are in very good agreement in each case, showing that our metallicities  
are on the same scale as that of their work.  Also the sample analyzed  here has a cluster in common with P15 (K44). For this cluster we found a metallicity value of -0.78 $\pm$ 0.03, which is 
in very good agreement with the value derived by P15 (-0.81 $\pm$ 0.04), based on completely independent data. This graphically confirms that our metallicities are also on the same scale as 
our previous work (including P09), which is expected considering that we have used the same telescope, instrument, instrumental configuration, methods and analysis.  Our current sample has no 
clusters in common with D21. However, we have a cluster in common (NGC151) with Dias et al. (submitted) which is based on observations taken with the same instrument and instrumental configuration 
as D21. Also, both works follow the same prescriptions and methods for the CaT metallicity determination. Therefore, the comparison with Dias et al. is an indirect comparison to D21.
Dias et al. (submitted) calculated a CaT metallicity of -0.75 $\pm$ 0.08 (private communication) in excellent agreement with the value found in this work. Consequently, we consider 
that our full cluster sample is homogeneous with metallicities on the same scale.

Although our sample is homogeneous in metallicity, it is necessary to note that it is heterogeneous in age, in the sense that the cluster data used in the literature to determine ages  are variable in instrument and quality (from precise HST data to less precise small telescope ground$-$based data).
Efforts to increase the cluster samples with homogeneous ages are being carried out by the VISCACHA survey \citep[][limiting magnitude V$\sim$24.5]{maia+19}, which aims at deriving ages of about half of the outer SMC clusters, among its multiple goals. The quality of the VISCACHA photometric data (obtained with the 4m telescope SOAR and its adaptive optics module SAM) and the precision of the methods used to determine both ages and other astrophysical parameters, have been shown in a series of publications with important results for the SMC study (\citealt{maia+19,santos+20}, D21).  Another important and complementary source of accurate photometry is the STEP survey \citep[][limiting magnitude $g \sim$ 24 ]{ripepi+14}. This survey, performed with the VLT Survey Telescope (VST), covers the SMC main body, the Bridge and part of the Magellanic Stream and allows the homogeneous determination of ages and structural parameters of a large cluster sample \citep{gatto2021}. 

We also emphasize that our cluster sample is not a magnitude or mass limited sample, nor is it complete in any sense. Therefore, a larger cluster sample will certainly reveal more details on the SMC chemical enrichment history.

\begin{table*}[!htb]
\caption{Extended cluster sample}             
\label{tab:cat_sample}      
\centering                          
\footnotesize
\begin{tabular}{l l l l l l l}        
\hline\hline                 
\noalign{\smallskip}  
  \multicolumn{1}{c}{Cluster} &
  \multicolumn{1}{c}{[Fe/H]} &
  \multicolumn{1}{c}{Ref} &
   \multicolumn{1}{c}{Age} &
   \multicolumn{1}{c}{Ref} &
   \multicolumn{1}{c}{$a$} &
   \multicolumn{1}{c}{$component$}  \\
     \multicolumn{1}{c}{} &
  \multicolumn{1}{c}{(dex)} &
  \multicolumn{1}{c}{} &
   \multicolumn{1}{c}{(Gyr)} &
   \multicolumn{1}{c}{} &
   \multicolumn{1}{c}{($^{\circ}$)} &
   \multicolumn{1}{c}{}\\
\noalign{\smallskip}
\hline 
\noalign{\smallskip}
BS95 121, OGLE-CL SMC 237                     &	-0.66 $\pm$	0.07 &	1  & 	2.8	 $\pm$ 0.5  & 5 & 1.4	& MB\\
B99, OGLE-CL SMC 122                          &	-0.84 $\pm$ 0.04 &	2  &	0.95	        & 6   & 1.2	& MB\\
H86-97, OGLE-CL SMC 43                        &	-0.71 $\pm$ 0.05 &	2  &	1.6	            & 6	 & 0.6	& MB\\
L17, K13, ESO 29-1                            &	-0.84 $\pm$	0.03 &  1	&	4.4	 $\pm$ 0.6	& 5 & 1.5	& MB\\
L19, OGLE-CL SMC 3	                          &	-0.87 $\pm$ 0.03 &  1	&	2.51 $\pm$ 0.1  & 7 & 1.5	& MB\\
L27, K21, OGLE-CL SMC 12	                  &	-1.14 $\pm$ 0.06 &  1	&	3.5	 $\pm$ 0.1	& 7 & 1.3	& MB\\
OGLE133                                       &	-0.80 $\pm$ 0.07 &	2	&	6.3	            & 6	 & 0.9	& MB\\
L1, ESO 28-8, OGLE-CL SMC 313                 &	-1.04 $\pm$ 0.03 &  2 	&   7.5  $\pm$ 0.5	& 9 & 5.0	& WH\\
L4, K1, ESO 28-15 	                          &	-1.08 $\pm$ 0.04 &  1	&	7.9	 $\pm$ 1.1	& 5 & 2.8	& WH\\
L5, ESO 28-16, OGLE-CL SMC 314	              &	-1.25 $\pm$ 0.05 &  1	&	3.7	 $\pm$ 0.5	& 5 & 3.0	& WH\\
L6, K4, ESO 28-17, OGLE-CL SMC 326            &	-1.24 $\pm$ 0.03 &  1	&	8.7	 $\pm$ 1.2	& 5 & 2.7	& WH\\
L7, K5, ESO 28-18, OGLE-CL SMC 324            &	-0.76 $\pm$ 0.06 &  1	&	1.6	 $\pm$ 0.2	& 5 & 2.5	& WH\\
L8, K3, ESO 28-19, OGLE-CL SMC 319            &	-0.85 $\pm$ 0.03 &  2	&	6.5	 $\pm$ 0.5	& 9 & 3.3	& WH\\
L9, K6, ESO 28-20, OGLE-CL SMC 332            & -0.63 $\pm$ 0.02 &	2  &	1.6	 $\pm$ 0.4	& 10 & 2.4	& WH\\
L12, K8                                       &	-0.70 $\pm$ 0.04 &	2  & 	1.3		        & 6	 & 2.4	& WH\\
L13, K9                                       &	-1.12 $\pm$ 0.05 &	2  &	0.4	            & 11 & 2.2	& WH\\
NGC121, L10, K2, ESO 50-12, OGLE-CL SMC 311   &	-1.19 $\pm$ 0.12 &	3	&	10.5 $\pm$ 0.5	& 8 & 4.8	& WH\\
HW86, OGLE-CL MBR 43                          &	-0.61 $\pm$ 0.06 &  1	&	1.4	 $\pm$ 0.2	& 5 & 6.8	& WB\\
L110, ESO 29-48, OGLE-CL SMC 292              &	-1.03 $\pm$ 0.05 &  1  &	7.6	 $\pm$ 1.0	& 5 & 4.9	& WB\\
L113, ESO 30-4, OGLE-CL MBR 47                &	-1.03 $\pm$ 0.04 &	2	&	3.98 $\pm$ 0.1	& 7 & 7.2	& WB\\
HW47                                          &	-0.92 $\pm$ 0.04 &	1  &	3.3	 $\pm$ 0.5	& 5 & 3.7	& SB\\
L58, K37                                      &	-0.79 $\pm$ 0.11 &	2	&	1.81 $\pm$ 0.24	& 12 & 2.7	& SB\\
L106, ESO 29-44, OGLE-CL SMC 296              &	-0.88 $\pm$ 0.06 &  1	&	2.0	 $\pm$ 0.3	& 5 & 7.8	& SB\\
L112, OGLE-CL SMC 298                         &	-1.08 $\pm$	0.07 &	2	&	5.1  $\pm$ 0.3	& 5 & 7.6	& SB\\
NGC 643, L111, ESO 29-50, OGLE-CL SMC 297     &	-0.82 $\pm$ 0.03 &	1  &	2.0	 $\pm$ 0.3	& 5 & 7.6  & SB\\
B168, OGLE-CL SMC 343                         & -1.08 $\pm$	0.06 &  4	&	6.60 $\pm$ 0.90	& 4 & 3.6	& NB\\
BS95 188, OGLE-CL SMC 302                     &	-0.94 $\pm$	0.06 &  4	&	1.82 $\pm$ 0.22	& 4 & 4.4	& NB\\
BS95 196, OGLE-CL MBR 36                      &	-0.89 $\pm$	0.04 &  4	&	3.89 $\pm$ 0.68	& 4 & 6.0	& NB\\
HW84, OGLE-CL SMC 305                         &	-0.91 $\pm$ 0.05 &	1  &	1.6	 $\pm$ 0.2	& 5 & 5.1  & NB\\
HW85                                          &	-0.82 $\pm$	0.06 &	4	&	1.74 $\pm$ 0.12 & 4 & 5.2	& NB\\
HW67, OGLE-CL SMC 335                         &	-0.72 $\pm$	0.04 &	2	&	2.7  $\pm$ 0.3	& 5 & 2.5	& NB\\
L100, ESO 51-27                               &	-0.89 $\pm$	0.06 &	4	&	3.16 $\pm$ 0.15 & 4 & 2.6	& NB\\
L102, IC1708, ESO 52-2, OGLE-CL SMC 342       &	-1.11 $\pm$	0.06 &	4	&	0.93 $\pm$ 0.16 & 4 & 3.3	& NB\\
L108, OGLE-CL SMC 300                         &	-1.05 $\pm$ 0.05 &	1  &	2.9	 $\pm$ 0.4	& 5 & 4.1  & NB\\
HW40                                          &	-0.78 $\pm$	0.05 &	2	&	2.5  $\pm$ 0.4	& 4  & 2.0	& CB\\
HW56, GLE-CL SMC 336                          &	-0.97 $\pm$	0.12 &	4	&	3.09 $\pm$ 0.22 & 4 & 2.4	& CB\\

\noalign{\smallskip}
\hline                                   
\end{tabular}\\
\textbf{References:} (1) P09, (2) P15, (3) DH98, (4) D21, (5) \citet{parisi+14}, (6) \citet{Rafelski+05},
(7) \citet{narloch+21}, (8) \citet{glatt+08a}, (9) \citet{glatt+08b}, (10) \citet{piatti+05}, (11) \citet{nayak+18}, (12) \citet{maia+19}. 
We only add NGC121 from DH98 because it is the only one not also analysed by P09, P15. The common clusters have consistent metallicities and we adopt P09,P15 values for homogeneity purposes.
\end{table*}

\subsection{Metallicity gradient}
\label{sec:met_grad}

MGs are important tools for analyzing the chemical evolution of galaxies and their dynamical history \citep{ho+15}. An examination of the MGs in nearby galaxies such as the MCs can help understand these processes in similar, more distant galaxies.  
Although there has been some discrepancy between spectroscopic and photometric studies regarding the existence of a MG in the SMC \citep{choudhury+20}, the latest research tends to suggest that field stars clearly present such a gradient. The spectroscopic work on red giant branch stars based on CaT metallicities (\citealt{carrera+08}, \citealt{dobbie+14b}, P10, P16) agrees on the existence of a clear MG in the SMC field. Although field star photometric studies \citep{piatti+12} and photometry of variable stars \citep{kapakos+11,haschke+12b} have not found evidence of a field MG, the photometric metallicity maps  created by \citet{choudhury+18} and \citet{choudhury+20} (using data from the Magellanic Cloud Photometric Survey (MCPS), the Optical Gravitational Lensing Experiment
(OGLE III) and the near-infrared VISTA Survey (VMC)), find evidence of metallicity trends in the SMC. In particular, \citet{choudhury+20} confirm that the SMC field MG gradients are radially asymmetric. 

\begin{figure}[!htb]
    \centering
    \includegraphics[width=\columnwidth]{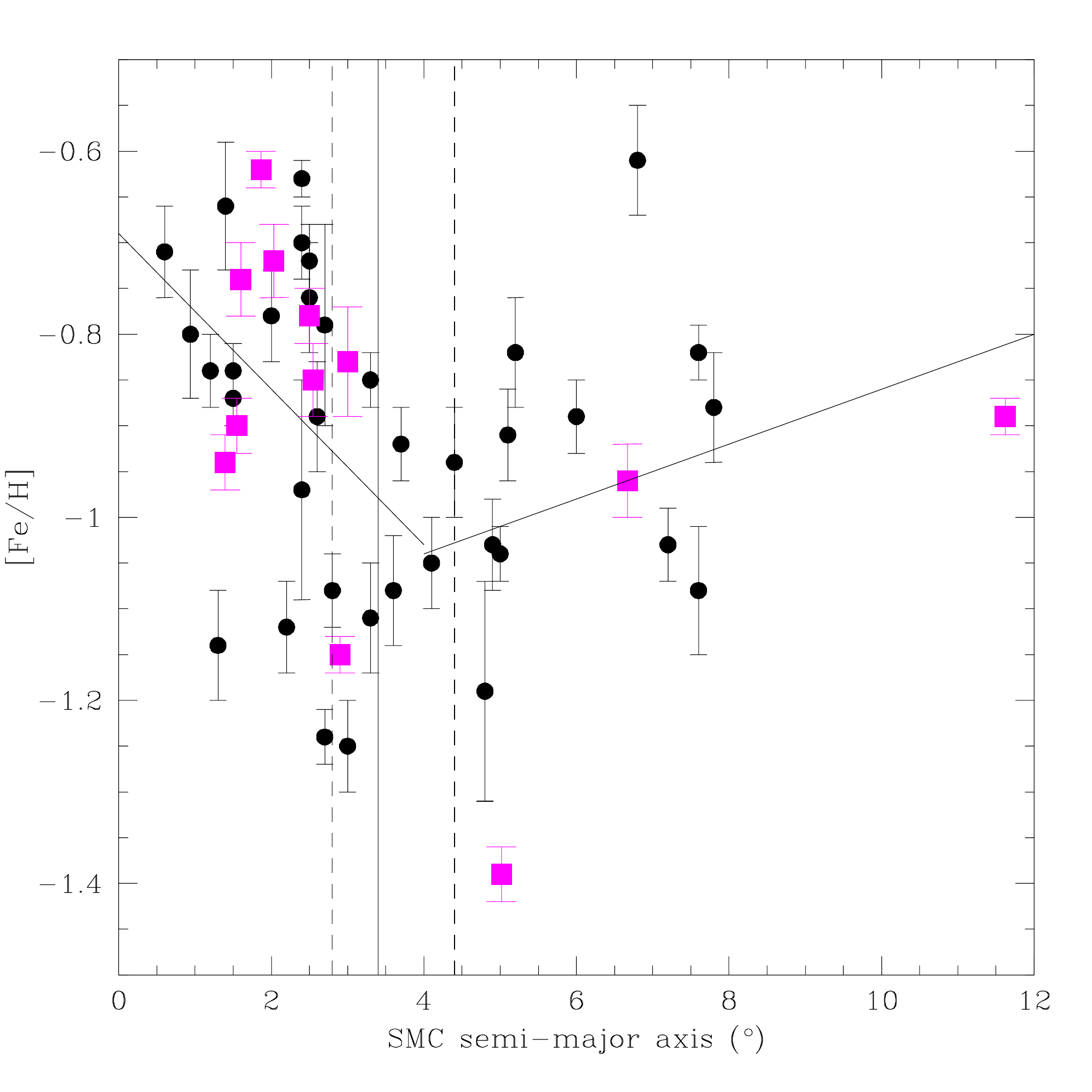}
    \caption{ Metallicity as a function of the semi-major axis $a$ for the full cluster sample. Black circles are clusters taken from the literature (Table \ref{tab:cat_sample}) and pink squares are clusters studied in this work (Table \ref{tab:results}). Vertical solid and dashed lines represent the SMC tidal radius and the errors from D21. The MG fits are shown for the inner and outer regions (solid lines).
        }
    \label{fig:mg}
\end{figure}

While there seems to be a general agreement in the literature that SMC field stars present a MG, the situation is not as clear when cluster samples are analyzed. P15, using a sample of 29 SMC clusters, showed that in the inner region of the galaxy ($a$ $<$ 4$^{\circ}$) it is not possible to find a statistically significant gradient, in contrast to what is observed in the outer part of the galaxy ($a$ $>$ 4$^{\circ}$) where the clusters present a behavior similar to that of field stars. In the external region of the SMC, both populations (clusters and fields) appear to present a positive MG with similar slopes. The difference in the behavior of clusters and field stars in the inner region of the SMC remains difficult to explain, 

Fig. \ref{fig:mg} shows the behavior of metallicity as a function of the semi-major axis $a$ for the complete sample. Pink and black symbols represent clusters included in Tables \ref{tab:results} and \ref{tab:cat_sample}, respectively. In this figure, the breakpoint found by D21 using the cluster radial density profile is marked with the solid vertical line  ($a=3.4^{\circ}$ $^{+1.0}_{-0.6}$).  \citet{massana+20} calculated that the tidal radius for the SMC is $\sim$ 4.5$^{\circ}$ based on the SMC and LMC masses and their relative distance. We adopted the value from D21 as the distance from the SMC center to divide it into what we call the internal and external regions because such distance is based on star clusters and is consistent with an independent calculation of the SMC tidal radius. In Fig. \ref{fig:mg}, it can be clearly seen that with this extended sample,  and using the projected distances adopted in this work, the metallicity spread is even larger in the internal region of the SMC, with respect to P15. 

We fit straight lines to the inner and outer regions, dividing at 3.4$^{\circ}$ (solid lines in Fig. \ref{fig:mg}) and we found values for the MG of -0.08 $\pm$ 0.04 dex deg$^{-1}$ and 0.03 $\pm$ 0.02 dex deg$^{-1}$, respectively. 
Clusters were equally weighted in making the fits. These values and their corresponding errors are consistent with the absence of a MG in the outer region.
Although the MG value obtained in the inner region is in agreement with that obtained for the field stars,  -0.075 $\pm$ 0.011 dex deg$^{-1}$ \citep{dobbie+14b} and -0.08 $\pm$ 0.02 dex deg$^{-1}$ (P16), it has an error of 50\% due to the large range in metallicities ($>$ 0.6 dex) that the clusters cover in that region. 

P15 drew attention to a possible inversion of the metallicity gradient (an aspect that they named the V-shape) in the external region of the SMC (beyond 4$^{\circ}$), which can also be seen in the compiled sample of clusters from the catalog of \citet{bica+20}. The V-shape is even more evident in SMC field stars studied with CaT (P16) but the field photometric study of \citet{choudhury+20} found that the metallicity  rises to an almost constant value of $-$0.93 dex from $\sim$ 3.5$^{\circ}$ to 4$^{\circ}$ (it is necessary to take into account that the definition of $a$ used by \citealt{choudhury+20} is not exactly the same as in this work). Significantly, the farthest  cluster in our sample (L116) belongs to the Southern Bridge and has a metallicity of -0.89, which is similar to the approximately constant metallicity in the outer regions found by \citet{choudhury+20}. If we consider the full cluster sample as in Fig. \ref{fig:mg}, the MG in the outer region appears to become flat, with a high dispersion, and the V-shape appears to be slightly diluted.  

\begin{figure}[!htb]
    \centering
    \includegraphics[width=\columnwidth]{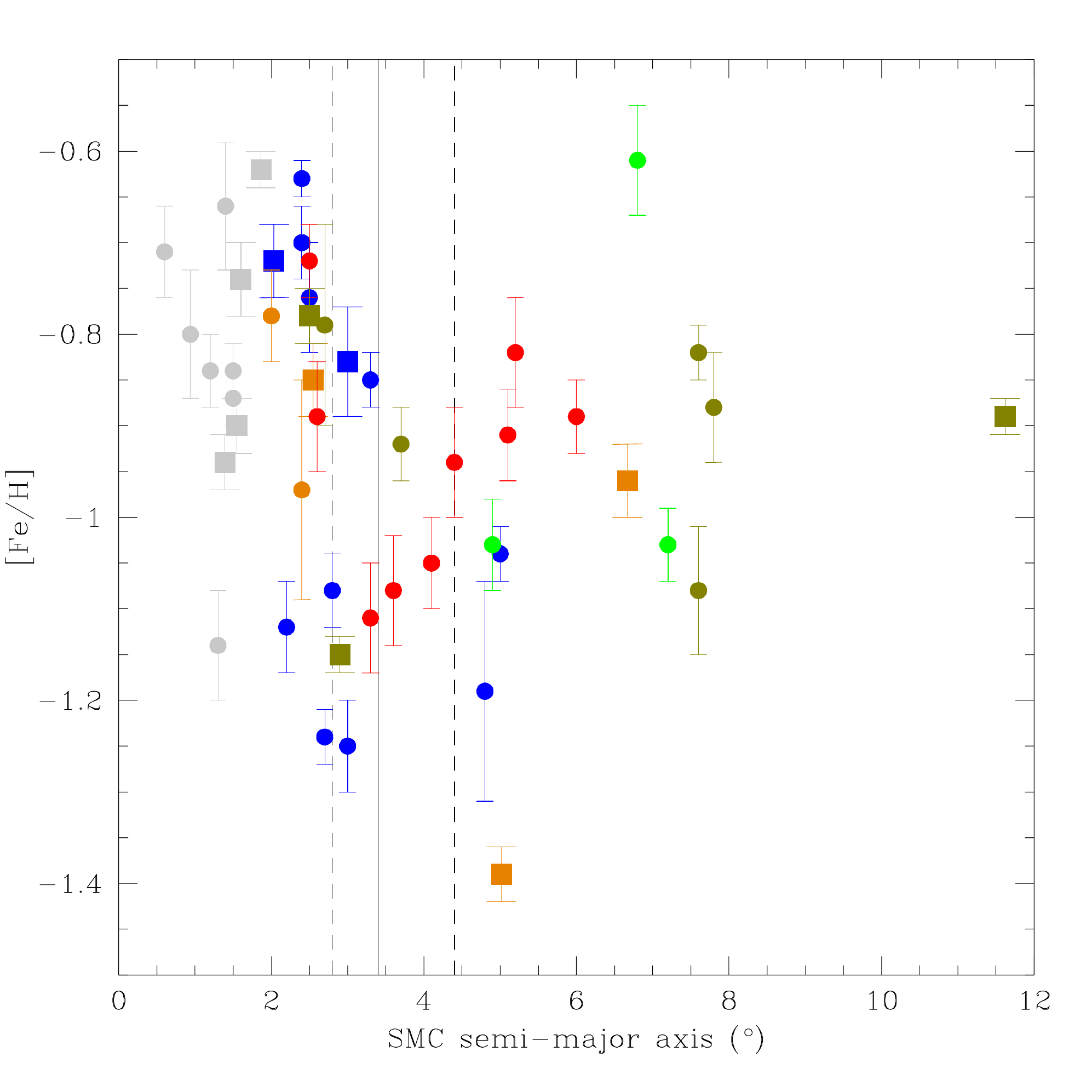}
    \caption{The same as in Fig. \ref{fig:mg} but symbols colored according to the adopted classification D14, D16b and D21: grey, orange, blue, green, red and brown symbols represent clusters belonging to the Main Body, Counter-Bridge, West Halo, Wing/Bridge, Northern Bridge and Southern Bridge, respectively. Squares are the clusters studied in this paper and circles represent our additional cluster sample. 
    }
    \label{fig:mgD}
\end{figure}

In Fig. \ref{fig:mgD} we plot the same data as in Fig. \ref{fig:mg} but use symbols with different colors according to the definition adopted in this work (see caption of Fig. \ref{fig:mgD}). One of the most conspicuous aspects of this figure is the fact that the only group of clusters that presents a particularly well-defined gradient or traces the V-shape in a clear way are those belonging to the Northern Bridge. It is  interesting to note, however, that all components show a minimum in metallicity as well as metallicity dispersion near the tidal radius ($\sim$ 0.2 dex) and that the dispersion 
grows considerably as we move away from this radius ($\sim$ 0.6 dex), both in the internal and external regions. Although the number of clusters in the Northern Bridge region present in our sample is still small, one gets the impression that it is the only group that shows a V-shape, with a vertex approximately coincident with the tidal radius, and an inversion of the gradient in the external region. D16b and  \citet{bica+20} argue that the V-shape is intrinsic to the Wing/Bridge region. Actually, the analysis that D16b carry out on a sample of clusters belonging to the West Halo does not show such an inversion, although the number of clusters in their West Halo sample beyond 4$^{\circ}$ is very small.  

\begin{figure}[!htb]
    \centering
    \includegraphics[width=\columnwidth]{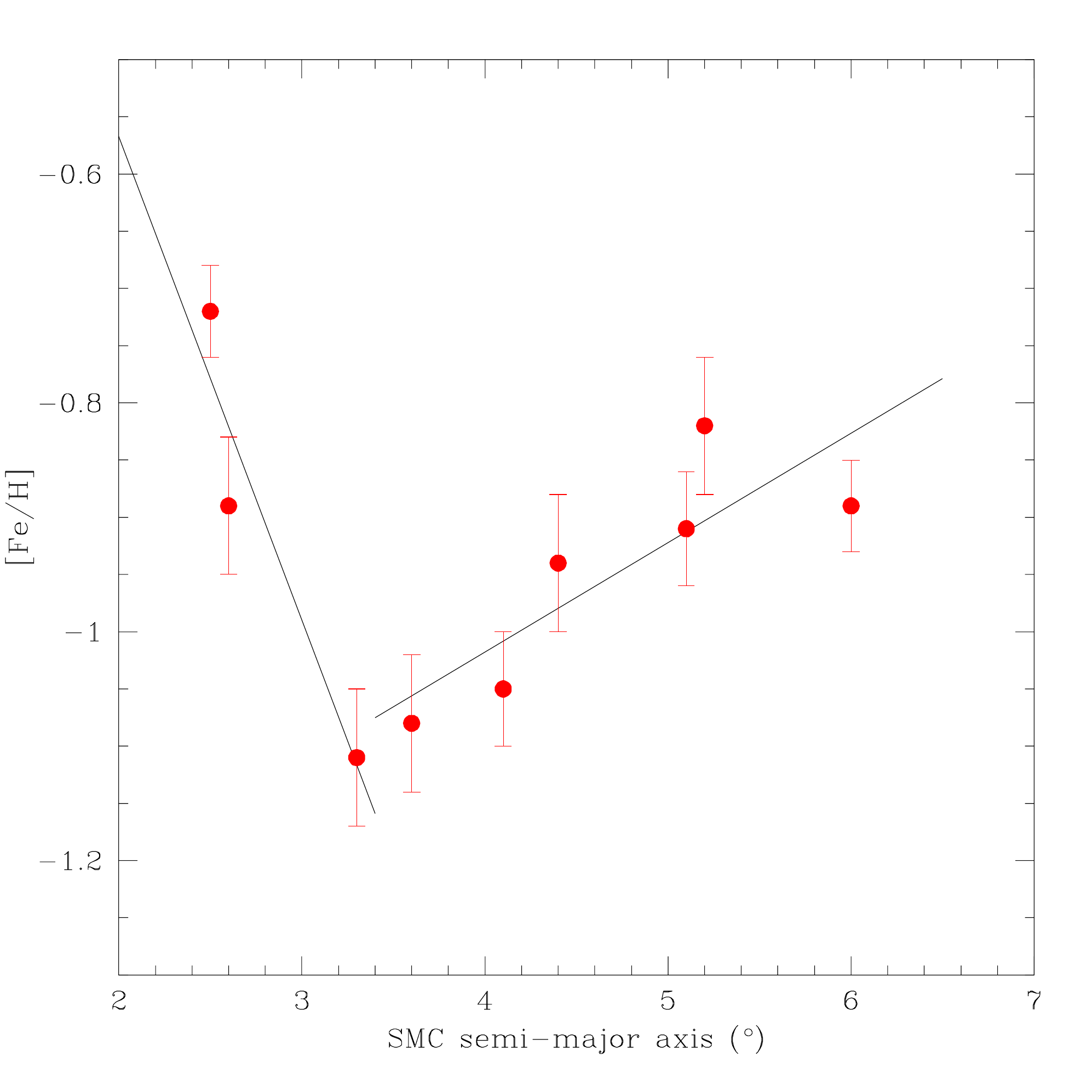}
    \caption{The same as in Fig. \ref{fig:mgD} but only for Northern Bridge clusters. Solid lines show the data fits inside and outside the tidal radius.  Note that the radial range is smaller than that shown in Figs. \ref{fig:mg} and \ref{fig:mgD}.
    }
    \label{fig:mgNB}
\end{figure}

Clusters belonging to the Northern Bridge classification have been graphed separately in  Fig. \ref{fig:mgNB},  showing a different radial range than in Figs. \ref{fig:mg} and \ref{fig:mgD} for a better visualization. The fits corresponding to the regions internal and external to the tidal radius are shown with solid lines in the figure. We found values of -0.42 $\pm$ 0.15 dex deg$^{-1}$ and 0.09 $\pm$ 0.03 dex deg$^{-1}$ for the inner and  outer region, respectively. The fit corresponding to the internal region in the Northern Bridge is not well constrained due to the low number of clusters. However, the external region presents a clear and much more robust indication of a positive gradient, which would be interesting to further investigate  with a larger sample. Mergers, interactions and radial migration could flatten or invert the metallicity radial gradient \citep[e.g.][]{tissera+16}.

As was already mentioned, D16b analyzed  the chemical properties of the West Halo. They determined photometrically the ages and metallicities of 9 clusters belonging to the West Halo, and supplemented their sample with 13 other clusters in the same region studied by other authors. Using that extended sample, they argued in favor of the existence not only of an MG, but also of an AG. The sample that these authors analyze includes metallicities determined with a variety of techniques and substantial differences in the quality of the data used in the compiled works, and thus their results are unfortunately plagued by inhomogeneity. For this reason, the  MG results from D16b are based on a sample of clusters with metallicities determined in an inhomogeneous way. Our sample includes 12 West Halo clusters, two in common with D16b: NGC152 and K8. D16b found metallicity value of -0.87 $\pm$ 0.07 for NGC152, close to the value derived in this work (-0.72 $\pm$ 0.04), but in the case of K8, D16b found a substantially  more metal-poor value (-1.12 $\pm$ 0.15) than in P15 (-0.70 $\pm$ 0.04). 

\begin{figure}[!htb]
    \centering
    \includegraphics[width=\columnwidth]{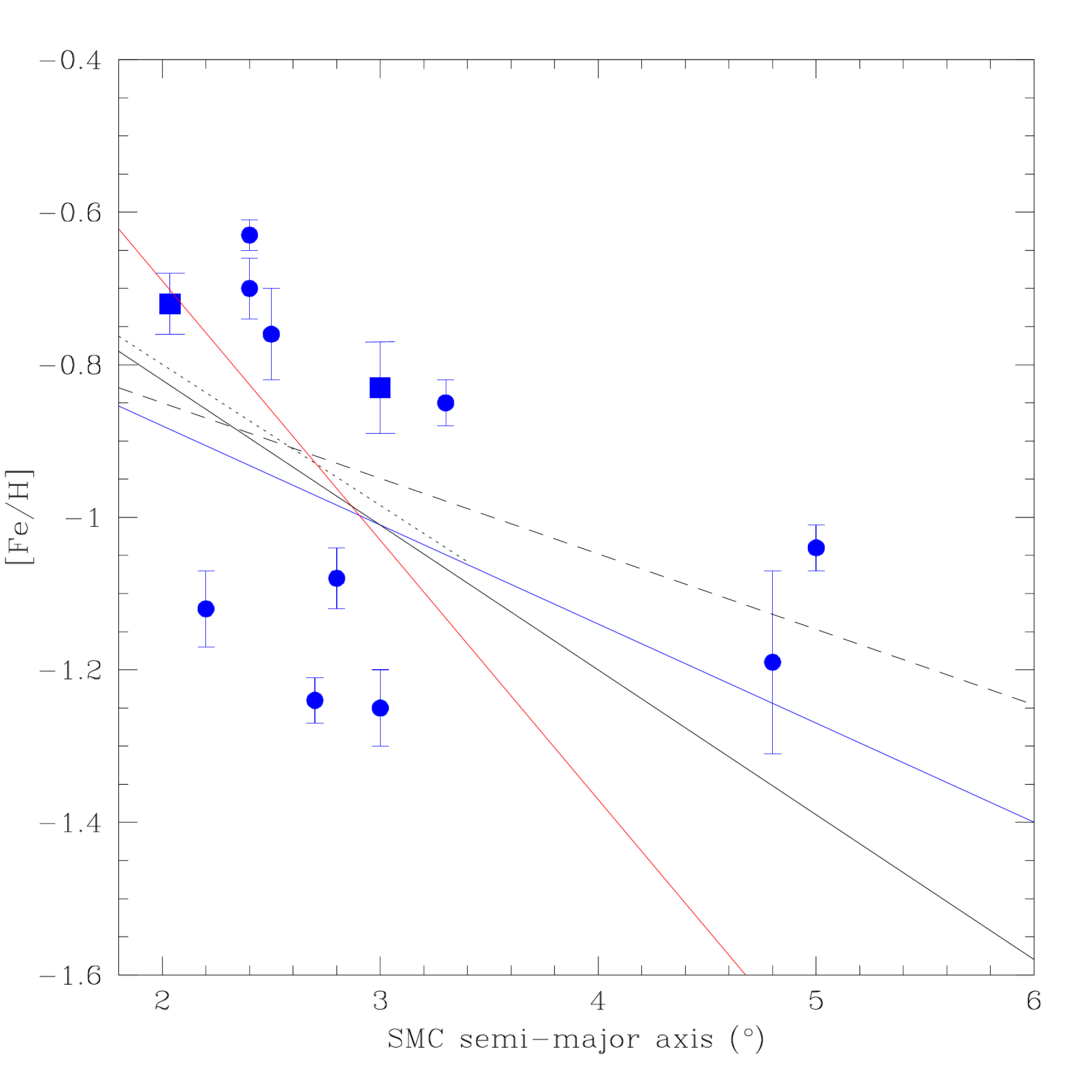}
    \caption{The same as in Fig. \ref{fig:mgD} but only for the West Halo clusters in our sample. Squares are clusters studied in the present work and circles are those taken from the literarure. Solid lines are the fits from D16b and dashed and dotted lines are the fits to our data. Note that the radial range is smaller than that shown in Figs. \ref{fig:mg} and \ref{fig:mgD}.
    }
    \label{fig:mgWH}
\end{figure}

\begin{figure}[!htb]
    \centering
    \includegraphics[width=\columnwidth]{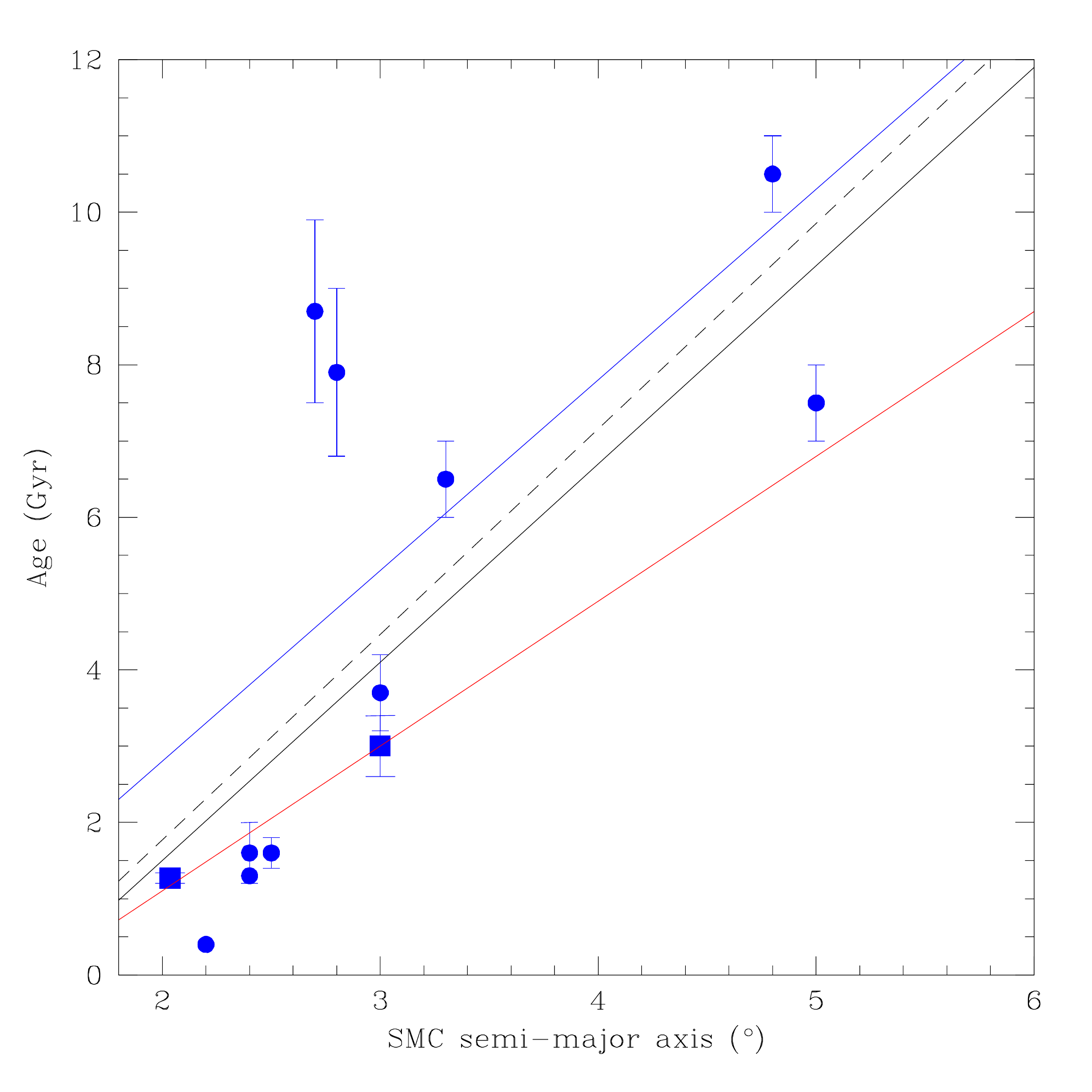}
    \caption{Age as a function of the semi-major axis $a$ for the West Halo clusters in our sample. The symbols are the same as in Fig. \ref{fig:mgWH}. The solid lines are the fits from D16b and the dashed line shows the fit to our data.
    }
    \label{fig:agWH}
\end{figure}

In order to analyze the existence of gradients in the West Halo with a sample of clusters observed and studied in the same way, and having precise metallicities on the same scale, we plot in Figs. \ref{fig:mgWH} and \ref{fig:agWH}  our sample belonging to that region.  We compare our homogeneous data with the fits derived by D16b, which can be seen in the figures indicated by solid lines. They performed fits to three different samples: clusters collected from the literature (blue line, -0.13 $\pm$ 0.08 dex deg$^{-1}$, 2.5 $\pm$ 0.8 Gyr deg$^{-1}$), clusters analyzed in their paper (red line, -0.34 $\pm$ 0.21 dex deg$^{-1}$, 1.9 $\pm$ 0.6 Gyr deg$^{-1}$) and all the clusters together (black line -0.19 $\pm$ 0.09 dex deg$^{-1}$, 2.6 $\pm$ 0.6 Gyr deg$^{-1}$). The linear fit to our complete West Halo sample yields a value of -0.09 $\pm$ 0.07 dex deg$^{-1}$ (dashed line). This value is in agreement with the MG for field stars but with an error of 78\%. If we consider only clusters within the 3.4 degrees from the SMC center, we obtain a MG of -0.19 $\pm$ 0.43 dex deg$^{-1}$ (dotted line). Although the value of the slope in the inner region is compatible with that found by D16b for their entire sample, our determination has a very large error. The uncertainties of the MG in our West Halo sample are of course affected by the considerable metallicity dispersion in the inner region.

Regarding the AG, we observe in Fig. \ref{fig:agWH} that our West Halo sample shows a clear tendency for the clusters to be older as the distance to the galaxy center increases. The fit to our data gives a value for the AG of 2.7 $\pm$ 0.8 Gyr deg$^{-1}$, in very good agreement with that found by D16b for their full sample.

While the latest photometric work investigating the age and metallicity gradients in the SMC through its star cluster system (e.g., \citealt{narloch+21}) suggests that the more metal-rich clusters are concentrated in the inner regions of the galaxy, our spectroscopic data show that the situation is more complicated. Although there seems to be a tendency for the clusters to be more metal-poor as we move away from the center of the SMC and out to 3.4$^{\circ}$, the dispersion of metallicities in the inner region is quite large, decreasing the significance of the linear fit.\\

D16b defined the Main Body as the region with $a <$ 2$^{\circ}$ based on visual criteria. However, taking into account both the results found here and those by \citet{dias+21}, we consider that this region should be defined by a more astrophysical criterion such as the tidal radius.  According to our results, we consider is reasonable to redefine the Main Body region as the one contained within a ellipse with a semi-major axis of 3.4$^{\circ}$, using D21's tidal radius value as our criterion.
\subsection{Metallicity distribution}
\label{sec:met_distr}

The MD of a galaxy's stellar populations contains relevant information to understand different astrophysical processes related to its evolutionary history, such as the star formation history, gas flows and chemical enrichment \citep[e.g.,][for dwarf galaxies]{kirby+13,leaman+13,fukagama20}. A possible bimodality in the SMC's cluster MD was suggested by P15. Based on a sample of 35 clusters with homogeneously determined CaT metallicities, they found a probability of 86\% that the cluster MD is bimodal, with possible peaks at -1.1 and -0.8 dex. 
This is contrary to what other studies suggest \citep[e.g.,][]{bica+20}. This latter study, based on a compilation of a large number of SMC and Magellanic Bridge clusters, found that the MD is unimodal, with a peak between -0.8 and -1.0. On the other hand, several studies have shown that the field metallicity distribution is unimodal with a peak between $\sim$ -0.94 to  $\sim$ -1 dex (e.g.,  \citealt{carrera+08}, P10, \citealt{dobbie+14b}, P16, \citealt{choudhury+18,choudhury+20}).\\

\begin{figure}[!htb]
    \centering
    \includegraphics[width=\columnwidth]{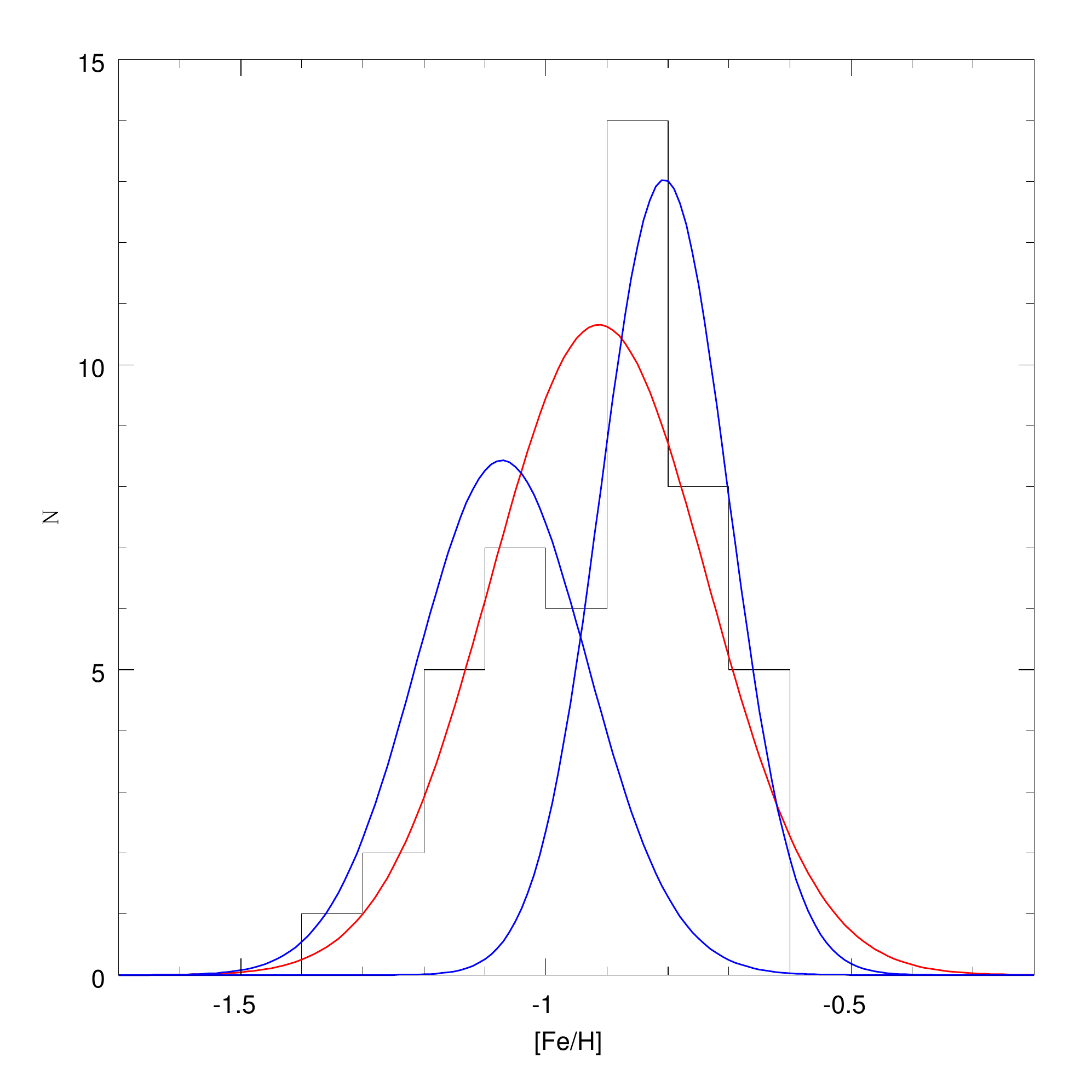}
    \caption{Metallicity distribution of our enlarged SMC cluster sample (Tables \ref{tab:results} and \ref{tab:cat_sample}). The fits derived from the application of the GMM algorithm are shown in red (unimodal fit) and blue (bimodal fit). The GMM method is independent of the bin size used for plotting the histogram. 
    }
    \label{fig:md}
\end{figure}

Using the sample analyzed in this work, we applied the Gaussian Mixture Model \citep[GMM,][]{muratov+10} in order to analyse the possible bimodality in the MD. The unimodal fit gives $\mu = $-0.913 and $\sigma =$0.178. The bimodal fit (heteroscedastic split) gives $\mu_1 =$ -0.806, $\mu_2 =$ -1.072, $\sigma_1 =$ 0.105 y $\sigma_2 =$ 0.140, with a p value of 0.476. This means that there is a 47.6\% probability of being wrong in rejecting unimodality. The parametric bootstrap gives a probability of 58.7\% that the MD is bimodal. The calculated separation of the peaks and the kurtosis value are 2.86 $\pm$ 0.89 and 0.476, respectively, compatible with a unimodal distribution. This means that with a larger cluster sample a bimodal MD is less significant than that found in P15. Our cluster MD, together with the unimodal (red line) and bimodal (blue line) fits are shown in Fig. \ref{fig:md}. Note that the GMM algorithm does not use the bin size of the histogram to make the probability calculations.

\subsection{Age-metallicity relation}
\label{sec:amr}

The AMR is a potentially  very useful tool to analyze the chemical history of a galaxy, providing hints on possible chemical enrichment processes. Various efforts have been made in recent decades to try to establish the AMR of the SMC, using as tracers both star clusters and field stars. Various models of chemical evolution have been proposed in the literature to explain the history of chemical evolution of the SMC. We found in our previous investigations, using a sample of 29 clusters (P09, P15), that there is no unique AMR in the SMC and that at a given age there is a dispersion of metallicities of 0.5 dex. This value cannot be explained by the errors involved in the calculation of metallicities, which are significantly lower. This dispersion of metallicities can also be observed in photometric studies of the SMC AMR (for example, \citealt{perren+17,narloch+21}).\\

\begin{figure}[!htb]
    \centering
    \includegraphics[width=\columnwidth]{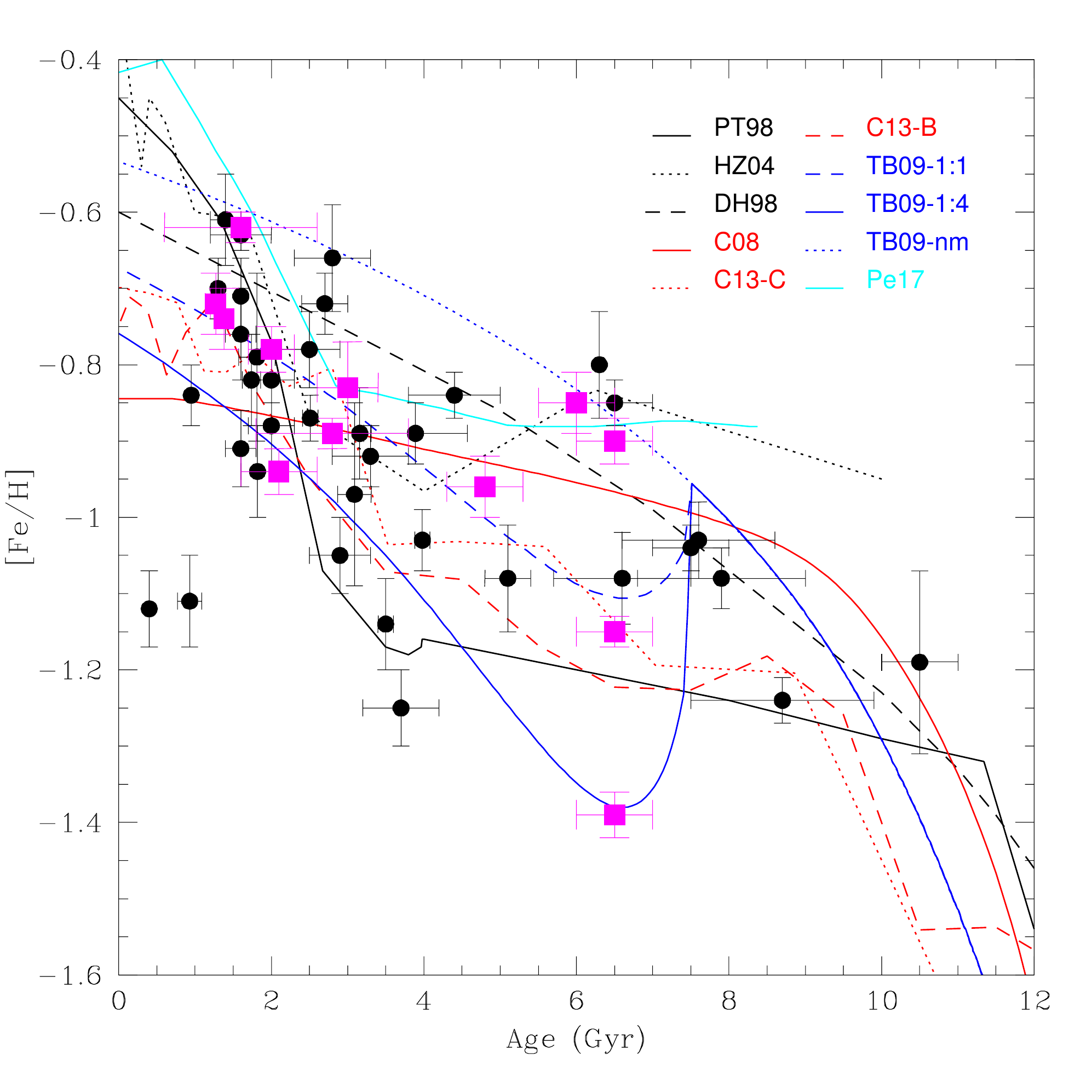}
    \caption{Metallicity as a function of age for the complete cluster sample. Magenta squares are the clusters analyzed in this paper. Black circles represent the additional sample of clusters with homogeneously derived spectroscopic metallicities. The observational data are compared with different models of chemical evolution available in the literature. The reference for each model is given in the inset. 
    }
    \label{fig:amr}
\end{figure}

We compare now in Fig. \ref{fig:amr} our enlarged cluster sample with the different models. The theoretical bursting model \citep[][PT98]{pagel+98} posits an initial star formation burst followed by a long period with no chemical enrichment (between 11 and 4 Gyr ago) and a more recent star formation burst that could have increased the metallicity in the SMC to its present value.  It is necessary to emphasize that the PT98 model  predicts basically no star formation between ages of $\sim$ 12 Gyr and $\sim$ 3$-$4 Gyr, which conflicts with the existence of clusters and field stars in that age range. An AMR model not only needs to fit the observational AMR, but it also needs to be consistent with the observed star formation history (SFH). 
The closed box model proposed by DH98, from the CaT analysis of six clusters distributed throughout the
spatial extent of the SMC, suggests a continuous and gradual chemical enrichment throughout the life of the SMC. The AMR from \citet{harris+04} (HZ04) is derived from the SFH analysis in 351 regions in the SMC across the central area of the main body (4$^{\circ}$ $\times$ 4.5$^{\circ}$). The AMR proposed by \citet{carrera+08} (C08)  comes from the CaT study of 350 red giant stars in 13 fields located between $\sim$ 1$^{\circ}$ and 4$^{\circ}$ from the SMC center. \citet{cignoni+13} proposed two AMR models (C13-C, C13-B) from the SFH of 4 fields, observed with the HST, located in the main body and the wing of the SMC, 0.5–2$^{\circ}$ from the galaxy center. The three models from \citet{tsujimoto+09} represent merger models with a mass radio of 1:1 (TB09-1:1) and 1:4 (TB09-1:4), and with no merger (TB08-nm). \citet[][hereafter Pe17]{perren+17} homogeneously analyze a sample of 89 SMC clusters, spatially distributed throughout a large area of the galaxy, using data in the Washington photometric system and the Automated Stellar Cluster Analysis (ASteCA) package \citep{perren+15}. They propose a model of chemical evolution with the metallicity decreasing towards older ages up to approximately 3 Gyr ago, similar to what is predicted by other models but moving towards higher metallicities. Their model resembles the AMRs proposed for field stars by HZ04 and \citet[][C13-C and C13-B]{cignoni+13} but shifted in metallicity to higher values.

In Fig. \ref{fig:amr}, we can see that our conclusions do not change substantially with this larger sample with respect to those reached in our previous work. For ages less than 4 Gyr, clusters appear to have undergone chemical enrichment similar to that predicted by the bursting model, with the exception of two clusters (K9 and IC1708). These two objects  have a considerably lower metallicity than predicted by any model for their ages. The cluster sample older than 4 Gyr presents a considerable dispersion of metallicities showing no agreement with any of the proposed chemical evolution models. Note that the addition of our most metal-poor cluster firms up the establishment of a metallicity spread at around 6.5 Gyr, now almost approaching 0.6 dex! The plot thickens!

\begin{figure}[!htb]
    \centering
    \includegraphics[width=\columnwidth]{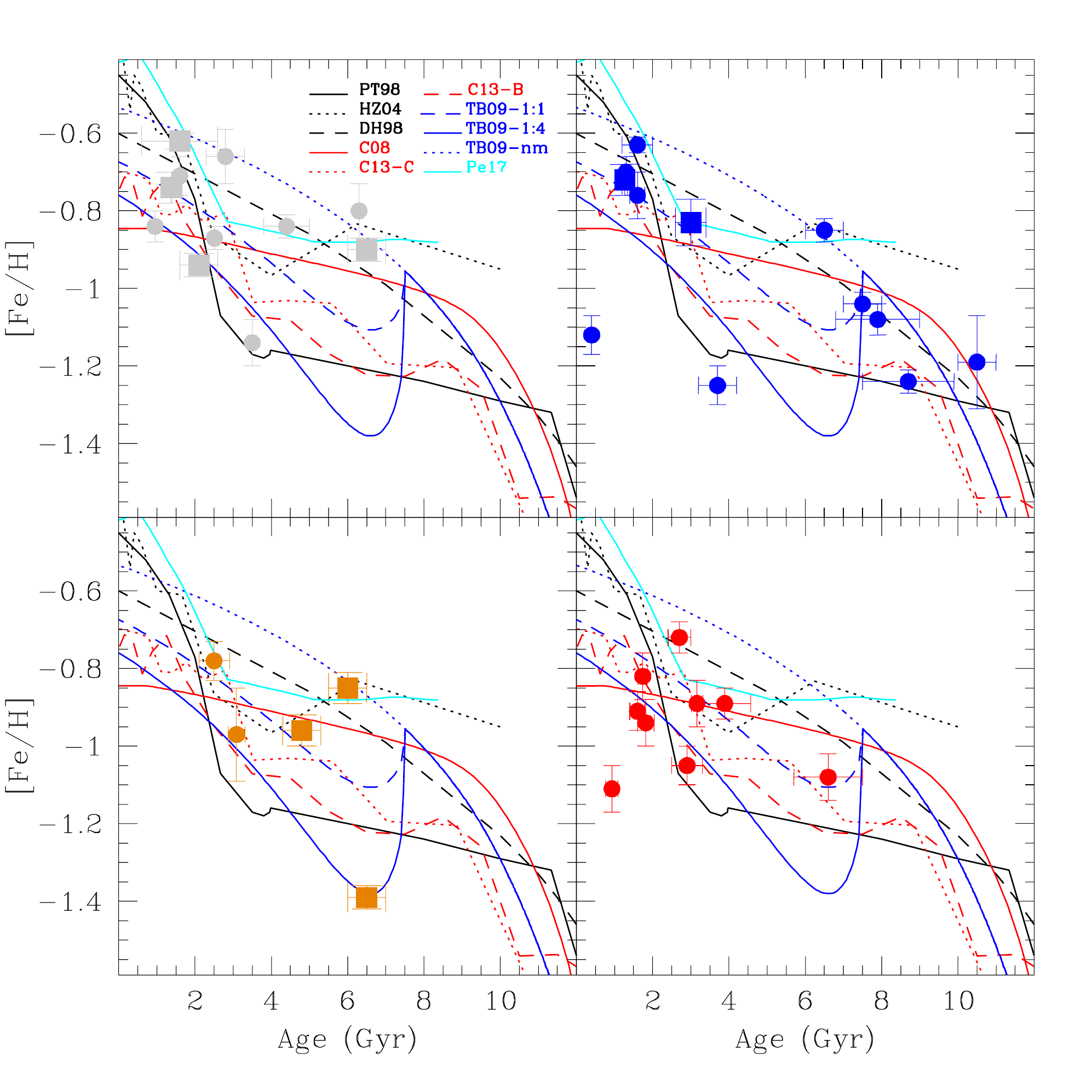}
    \caption{The same as in Fig.\ref{fig:amr} but with the cluster sample identified according to the classifications of D16b and D21. The color code is explained in Fig. \ref{fig:mgD} and in the text. 
    }
    \label{fig:amrD}
\end{figure}

\begin{figure}[!htb]
    \centering
    \includegraphics[width=\columnwidth]{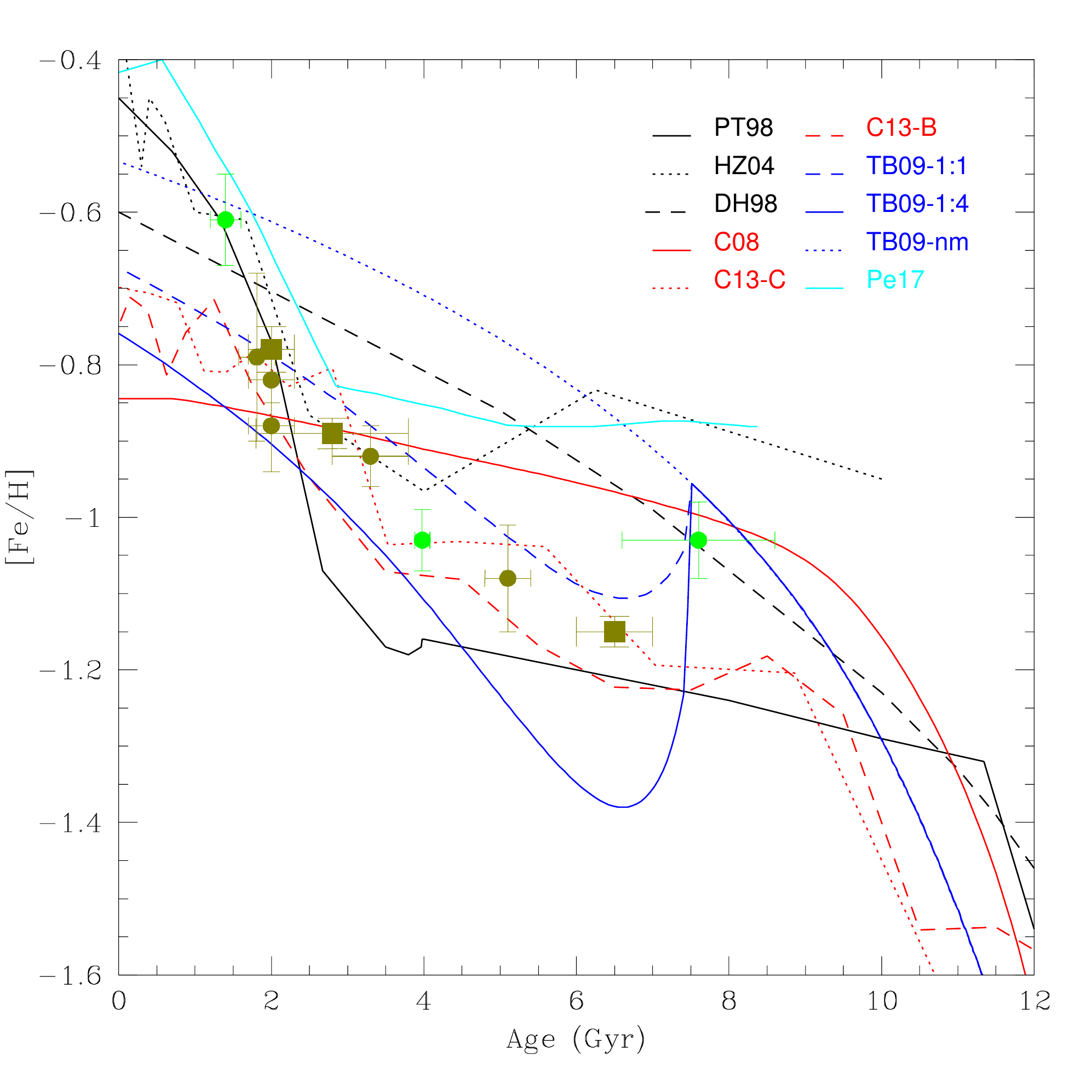}
    \caption{The same as in Fig. \ref{fig:amr} but for Wing/Bridge and Southern Bridge clusters. 
    }
    \label{fig:amrWB_SB}
\end{figure}

As we did in the previous section, we analyze the AMR by separating the cluster sample according to the classification proposed by D16b and D21, which can be seen in  Fig. \ref{fig:amrD}. The first thing that stands out from this figure is that the Main Body, Counter-Bridge and Northern Bridge would not seem to present a clear AMR, each covering a wide range of ages and metallicities. 
Additionally, the West Halo clusters for which D16b photometrically found an AMR compatible with the bursting model, appear to follow the closed box model (DH98) according to our data, with the exception 
of two points.
 The clusters belonging to the Wing/Bridge and Southern Bridge, whose AMR has been graphed separately in Fig. \ref{fig:amrWB_SB}, seem to share the same chemical evolution, although it is difficult to be conclusive due to the low number of Wing/Bridge and Southern Bridge clusters  present in our sample. Assuming that the SMC does have a unique AMR in both those regions, it would seem to follow the predictions of the bursting model but with metallicity values higher than those expected for clusters older than 3-4 Gyr. In fact, in these regions the model of chemical evolution in that SMC region appears to be an intermediate model between those of  \citet[][PT98]{pagel+98} and Pe17. These two models predict a similar chemical enrichment history, but displaced relative to each other in metallicity, mainly at intermediate ages before the possible burst of star formation 4 Gyr ago. A precisely intermediate model between
these two scenarios, which would seem to fit approximately the data, is that of \citet{tsujimoto+09} corresponding to a 1:1 merger. Alternatively, the two AMRs proposed by \citet{cignoni+13} reproduce the data well. Particularly interesting are the C13-B and C13-C models, which were fitted to the main body and wing using HST data.  \\

We trust that with a database of clusters with homogeneous metallicities and ages, and with the accuracy provided by the CaT technique and the VISCACHA and  STEP data, it will be possible to disentangle the AMR in a more precise way, especially in the intermediate age range, if indeed it can be disentangled. Also, the observation of a larger number of clusters in the Wing/Bridge and Southern Bridge regions would be necessary to corroborate the possible AMR they have in common.  On the other hand, the SMC may just be complicated and challenge simple assumptions.  With the exception of the TB09 models, the proposed AMRs assume that the SMC did not experience accretion and mergers, whereas its highly disturbed shape may also be the result of a merger \citep{bekki+08,tsujimoto+09,pieres+17}. If so, more metal-poor clusters at a given age might derive from a smaller, more metal-poor dIrr that merged with the early SMC, partly producing the cluster metallicity dispersion that we observe.  Further, recent work, such as that described in, for example,  \citet[][]{li+21a} and Li et al. (2021b, in prep), suggests that even in a dwarf galaxy like the SMC, the correlation length scale for abundance enrichment is still substantially smaller than the galaxy's size.  Consequently, Supernova enrichment products are not expected to have been homogeneously mixed through the entire SMC.  Moreover, \citet{nidever+20} found that the SMC has had a very low star formation efficiency, even in comparison with other less massive dwarf galaxies when comparing the [$\alpha$/Fe] vs. [Fe/H] trends assuming that the efficiency should increase with galaxy mass. Therefore, the substantial spread in cluster abundances at fixed age is not unexpected. This does not rule out, however, the possibility of infall (and incomplete mixing) of pristine (or at least low abundance) gas, potentially connected to the SMC/LMC interaction history, contributing to the range in cluster abundances at fixed age.\\

Finally, it is necessary to bear in mind that the definition of D16b is an initial classification based on the projected position of the clusters without considering any information on the dynamics of the clusters. We need to know the cluster orbits in order to analyze possible superposition of different timescales in the same region. So far, no SMC cluster orbits have been calculated, but they will be part of a future work. The full kinematics of the SMC clusters has only recently started to be mapped (e.g. D21, \citealt{piatti21}, Dias et al. (2021, submitted).

\section{Summary and conclusions}
\label{sec:summary}

We present radial velocites (RVs) and Calcium Triplet metallicities (CaT) for a large sample of red giant stars in 12 SMC clusters. We derived mean cluster RVs and metallicities with a mean error of 2.1 km s$^{-1}$ and 0.03 dex, respectively. Using this information, together with that available in the literature for another 36 clusters with CaT metallicities derived homogeneously, we analyzed the metallicity gradient, the metallicity distribution and the age-metallicity relation in the SMC. Clusters of the final sample are distributed in an area of $\sim$ 70 deg$^2$ with 0.4 Gyr $\leq$ Age $\leq$ 10.5 Gyr. This is the largest sample of spectroscopically analyzed SMC clusters available to date. Following the ideas of  \citet{dias+16b} and \citet{dias+21}, we divided the sample in six groups: Main Body, Counter-Bridge, Wing/Bridge, Northern Bridge, Southern Bridge and West Halo. Also we adopted the value of the SMC breakpoint derived by \citet{dias+21} (3.4$^{\circ}$ $^{+1.0}_{-0.6}$) 
to divide the galaxy into what we call the inner and outer regions. We can summarize our results as follows:

\begin{itemize}
    \item We confirm that in the inner region (inside 3.4$^{\circ}$) the SMC clusters present a considerable dispersion of metallicity ($\sim$ 0.6 dex). 
    \item Clusters in the inner region exhibit a metallicity gradient with a value compatible with that shown by field stars but with an error of 50\% due to the large dispersion of metallicities. On the other hand, our data show that the outer region does not present a significant MG.
    \item Concerning the components suggested by Dias et al., we observed that only the clusters belonging to the Northern Bridge appear to trace a V-Shape metallicity gradient, showing a clear inversion near the tidal radius of the SMC. However, all the groups  present a minimum in the metallicities at a distance from the center of the galaxy that coincides with the tidal radius proposed by \citet{dias+21}.
    \item Our sample of West Halo clusters shows a clear age gradient, in agreement with \citet{dias+16b}. Regarding the MG we found a value compatible with the one derived for field stars \citep[e.g.,][]{parisi+16,dobbie+14b} and with the MG derived by \citet{dias+16b} for the West Halo, depending on whether we consider all of our West halo clusters or only those located in the inner region.
    \item  The difference in the behavior of the metallicity gradient inside and outside the tidal radius, as well as the differences found analyzing the different groups defined by \citet{dias+16b} and \citet{dias+21}, suggests that the chemical history of the SMC strongly depends on its dynamical history, as was previously emphasized by those authors. Also, according to these results, we propose to extend the Main Body region out to 3.4$^{\circ}$.
    \item  The application of the Gaussian Mixture model \citep{muratov+10} to the metallicity distribution of our entire cluster sample points to the probability that our cluster metallicity distribution is much less bimodal than  was previously found by \citet{parisi+15}.
    \item With respect to the age-metallicity relation, clusters younger than 4 Gyr appear to follow the bursting model of \citet{pagel+98}. For intermediate ages, no model adequately reproduces the data and the dispersion of metallicities becomes even more evident.  The development of theoretical models that reproduce the observations and also satisfy the SMC cluster and field star SFHs are essential for a complete understanding of the chemical evolution of this galaxy. We hope that our results will be an inspiration in that sense.
    \item  Clusters belonging to the Wing/Bridge and Southern Bridge exhibit a well-defined age-metallicity relation with relatively little scatter in abundance at fixed age compared to the other regions. Although the sample is small, both groups appear to share a common chemical enrichment history. Our data also suggest that the West Halo clusters could follow the closed box model proposed by \citet{dacosta+98}.
    \item The lack of a clear age-metallicity relation for the SMC as a whole, and the large spread of metallicities at any given age, could indicate that mergers, including gas infall, played a role during its history, in addition to the interaction with the LMC.
\end{itemize}

\begin{acknowledgements}

This work is based on observations collected at the European Southern Observatory, Chile, under Program 075.B-0548 and 073.B-0488. 
This work presents results from the European Space Agency (ESA) space mission Gaia. Gaia data are being processed by the Gaia Data Processing 
and Analysis Consortium (DPAC). Funding for the DPAC is provided by national institutions, in particular the institutions participating in the 
Gaia MultiLateral Agreement (MLA). The Gaia mission website is https://www.cosmos.esa.int/gaia. The Gaia archive website is https://archives.esac.esa.int/gaia.
We thank the referee for comments that helped to improve this work. This research was partially supported by the Argentinian
institution SECYT (Universidad Nacional de Córdoba).
D.G. gratefully acknowledges support from the Chilean Centro de Excelencia en Astrof{\'\i}sica
y Tecnolog{\'\i}as Afines (CATA) BASAL grant AFB-170002.
D.G. also acknowledges financial support from the Direcci\'on de Investigaci\'on y Desarrollo de
la Universidad de La Serena through the Programa de Incentivo a la Investigaci\'on de
Acad\'emicos (PIA-DIDULS).

\end{acknowledgements}

   \bibliographystyle{aa} 
   \bibliography{bibliography} 

\appendix

\end{document}